\documentclass[lettersize,journal]{IEEEtran}
\usepackage{amsmath,amsfonts}
\usepackage{algorithm}
\usepackage{array}
\usepackage[caption=false,font=normalsize,labelfont=sf,textfont=sf]{subfig}
\usepackage{textcomp}
\usepackage{stfloats}
\usepackage{url}
\usepackage{verbatim}
\usepackage{graphicx}
\usepackage{cite}
\usepackage{todonotes}
\usepackage{xcolor}
\usepackage{booktabs}
\usepackage{multirow}
\usepackage{algorithmicx}
\usepackage{algpseudocode}

\begin{document}

\title{T2S: A Rehearsal-Based Approach for Extraction-Resistant Model Watermarking}

\author{
    Jian-Ping Mei\IEEEauthorrefmark{1}\thanks{Corresponding author: J.-P. Mei}, 
    Weibin Zhang\IEEEauthorrefmark{1}, 
    Ao Yao\IEEEauthorrefmark{1}, 
    Tiantian Zhu\IEEEauthorrefmark{1}
    Jie Xiao\IEEEauthorrefmark{1}
    \\
    \IEEEauthorrefmark{1}College of Computer Science and Technology, Zhejiang University of Technology, Hangzhou, 310023 China \\
    \{jpmei,ttzhu,xiaojieqj\}@zjut.edu.cn, zhangweibin30@gmail.com, aoyao6697@gmail.com

\thanks{This paper was supported by the National Natural
Science Foundation of China (Grant No.: 62276234), and the Zhejiang Provincial Natural Science Foundation (Grant No.: ZCLZ24F0202).}

}

\markboth{Journal of \LaTeX\ Class Files,~Vol.~14, No.~8, August~2021}%
{Shell \MakeLowercase{\textit{et al.}}: A Sample Article Using IEEEtran.cls for IEEE Journals}


\maketitle

\begin{abstract}
Model watermarking safeguards AI model intellectual property by embedding distinctive knowledge that induces unique behavioral signatures. The primary technical challenge lies in ensuring watermark robustness against various post-processing attacks on the watermarked model. Model extraction attacks emerge as the most severe threat, where adversaries exploit prediction outputs to train surrogate models that illegally replicate the original model's functionality.
In this work, we propose a rehearsal-based watermark embedding framework to enhance the robustness of model watermarks against model extraction attacks. By simulating the extraction process, our method leverages the loss of a \textit{simulated stolen model} on a trigger set as a training signal to fine-tune the watermark knowledge within the target model. This fine-tuning step encourages the watermark to be embedded in a way that boosts transferability, thereby increasing its chances of persisting and remaining detectable in stolen models.
Comprehensive experiments conducted under diverse settings demonstrate that the proposed method significantly improves the robustness of model watermarks against both model extraction and subsequent watermark removal attacks.

\end{abstract}

\begin{IEEEkeywords}
Model watermarking, Rehearsal-based embedding, Model extraction attack, Watermark removal attack 
\end{IEEEkeywords}


\section{Introduction}

\IEEEPARstart{D}{eep} neural networks (DNNs)  based vision recognition have emerged as foundational technologies in artificial intelligence, powering critical applications such as autonomous driving \cite{Renz0AS25}, industrial robotic vision \cite{DF25}, and computer-aided diagnosis \cite{zhou2024skinGPT4}. Training DNNs typically demands significant resources, including large-scale or proprietary datasets, substantial computational power, and considerable human effort, rendering these models valuable assets. 
Many AI companies offer machine learning services via APIs to generate profit from pay-per-use systems. However, malicious users may replicate the functionality of a target model using model extraction techniques. For example, they can query a classification model with surrogate or synthesized data and train their own models based on the returned label information\cite{orekondy19knock, jagielski20, kariyappa21datafree}.
Such attacks violate the model owner’s intellectual property rights, leading to unfair competition and potentially exposing privacy and security issues \cite{shokri17}.

 Watermarking has emerged as a representative passive defense mechanism to safeguard against model extraction attacks \cite{uchida17}. It involves embedding specific knowledge into the model, which enables the model to exhibit unique behaviors, such as producing distinctive outputs for predetermined inputs. These predefined input-output pairs are known as trigger samples. When a suspicious model is believed to be a functional copy trained with model extraction techniques, the owner of the original model can prove ownership by detecting the watermark in the suspect model with a trigger set. 


Early methods employed out-of-distribution (OOD) images or images stamped with unrelated patches as trigger inputs \cite{zhang18}. While such watermarks could be successfully embedded with a small negative impact on the performance of the target model, retaining watermarks within stolen models obtained through model extraction is quite challenging \cite{jia21,lukas22, adi18}. This difficulty arises primarily because the model extraction process prioritizes the replication of the target model's acquired knowledge particularly with respect to the classification task. If the model treats watermark information as an independent feature detached from those used for the classification task, the neurons activated by watermark triggers may decouple from those handling normal inputs, rendering the watermark irretrievable during model extraction.

Some latter studies proposed to improve watermark retention in stolen models by increasing the distribution closeness between the watermark trigger samples and the training data, either in the input space or feature space \cite{jia21,lv24}. Another approach, called \emph{SSW} \cite{tan23}, proposed to watermark the target model with trigger samples fine-tuned using a simulated stolen model (SSM).
In this paper, we propose a new rehearsal-based watermark fine-tuning framework for robust model watermarking. While \emph{SSW} relies on a fine-tuned trigger set as an intermediary to implicitly reflect watermark retention of the SSM back to the target model, our approach directly back-propagates the feedback from the SSM to the target model. This new formulation leads to much higher watermark detection rates in stolen models together with trivial false positive rates without the help of any negative models. The watermarked stolen model is also resilient to subsequent watermark removal processes. Our main contributions include:

\begin{itemize}
\item Propose a novel rehearsal-based framework that enables the target model to fine-tune its initially learned watermark knowledge directly with the watermarking loss of simulated stolen model.
\item Instantiate the proposed watermark fine-tuning framework with three different types of trigger sets to show its generality and versatility.
\item Provide comprehensive experimental results to demonstrate the strong resistance of our approach to various model extraction attacks and subsequent watermark removal operations.
\end{itemize}

In the following sections, we first review related work in model extraction attack and DNN model watermarking in Section \ref{sec:related work}. We then present the details of the proposed approach in Section \ref{sec:proposed method}, and provide experimental results in \ref{sec:experiment}. Finally, we conclude in Section \ref{sec:conclude}

\section{Related Work}
\label{sec:related work}

\subsection{Model extraction attack}
Model extraction attacks \cite{tramer16} aim to replicate a target model's functionality without authorization. Typically, attackers query the black-box model through APIs to get the predictions and then train a substitute model using these input-output pairs. The query data can either be an appropriate proxy dataset \cite{orekondy19knock} or synthetic samples generated by generative models \cite{kariyappa21datafree,truong21,sanyal22}. These model extraction methods can produce stolen models comparable to the target model in performance, given a sufficient number of queries.



Defenses against model extraction can be categorized into active and passive approaches.  Active defenses modify the model predictions in response to queries to disrupt the training of stolen models \cite{orekondy19pp,kariyappa21ensemble,kariyappa20,chen20,zhang23}.
While active defenses offer timely protection, they usually depend on detecting malicious queries by assuming that the attacker's queries differ more significantly from the legitimate user queries to the target model’s training data \cite{juuti19, kariyappa20,chen20,zhang23}.
However, in practice, identifying queries from malicious users is challenging, thereby limiting the effectiveness of active defenses.


\subsection{DNN watermarking}
Model watermarking is a passive defense mechanism designed to prove model ownership after theft by detecting embedded watermarks. Since while-box approaches inspired by digital watermarking have limited applicability due to the requirement of accessing internal information and responses from suspicious models during watermark extraction in the verification phase \cite{uchida2017embedding, nagai2018digital,chen2019deepmarks,darvish2019deepsigns,tartaglione2021delving}, we focus on black-box model watermarking in this paper.
The watermark is encoded through distinct model responses to a private trigger set, consisting of carefully crafted input-output pairs known exclusively to the model owner. Only black-box access is required for watermark verification.

The existing black box model watermarking techniques are mainly divided into backdoor-based and adversarial sample-based approaches. Backdoor-based methods leverage model backdoors as watermarks by crafting unique trigger samples, ranging from manually designed abstract patterns \cite{adi18} to noise- or text-overlaid training images, out-of-distribution (OOD) samples \cite{zhang18}, or optimized small perturbations, which are assigned to a target class to form a trigger set. This set is then combined with normal training data to embed the watermark during model training. Meanwhile, adversarial sample-based methods focus on exploiting adversarial samples or synthetic samples near the decision boundaries of the target model as trigger samples.


 However, successfully embedding a watermark into a model represents only the first step. The more critical challenge lies in ensuring that the watermark remains effective after undergoing various model post-processing, such as fine-tuning, model compression, and model extraction. 
To improve watermark retention under model extraction, several approaches have aimed to strengthen the coupling between the watermarking and the primary learning tasks while maintaining watermark uniqueness. In \cite{jia21}, an auxiliary loss is introduced to encourage trigger samples to intertwine with training data in the feature space. Other studies focused on trigger-set construction, enhancing correlation between the two tasks. In \cite{lv24}, trigger samples are generated by blending training samples from two classes, a method initially introduced for backdoor attacks. The idea of simulation-based trigger optimization was first explored for backdoor attacks in \cite{ge21} and later extended to model watermarking in \cite{tan23}.

Although model watermarking and backdoor attacks \cite{gu19} share conceptual similarities, they differ substantially in trigger flexibility and training control. Watermarking allows defenders to freely design trigger sets and typically have fully control over the training process, whereas backdoor attackers are constrained by trigger invisibility and the lack of access to the target model as well as its training for more practical scenarios. Moreover, watermark triggers can be reused for verification, while backdoor triggers typically operate only on unseen samples.


\section{Proposed approach}
\label{sec:proposed method}
\subsection{Threat Model}
\label{preliminaries}
\subsubsection{Defender}
When embedding a watermark into the target model, the defender need to consider the \textit{protection} together with \textit{utility} and \textit{uniqueness}. High \textit{utility} means that watermark embedding should have minimal impact on the model’s accuracy, ensuring that its primary task performance is not significantly degraded. The embedded watermark should exhibit uniqueness, being triggered only in the target model and its stolen versions, while not activating in other independently trained models. Moreover, the watermark must remain effective against model extraction and subsequent watermark removal techniques such as pruning, quantization and second extraction. In this paper, we consider the more practical black-box verification. This means that the defender evaluates the suspicious model's watermark triggering capability by querying it with trigger samples via API.


\subsubsection{Attacker}
We assume that the adversary is aware of the existence of a watermark embedded in the target model but has no knowledge of its specific form. The goal of the adversary is to replicate the functionality of the target model while attempting to remove the watermark. 
We consider the black-box model stealing scenario where the adversary is unaware of the target model's architecture. Attackers can only submit a limited number of queries with surrogate data (either real or synthetic) to train a stolen model that may architecturally differ from the target model.


\subsection{Motivation}
\begin{figure}[!t]
	\centering
	\subfloat[\emph{SSW} \cite{tan23}]{
        \includegraphics[width=0.5\linewidth,height=0.33\linewidth]{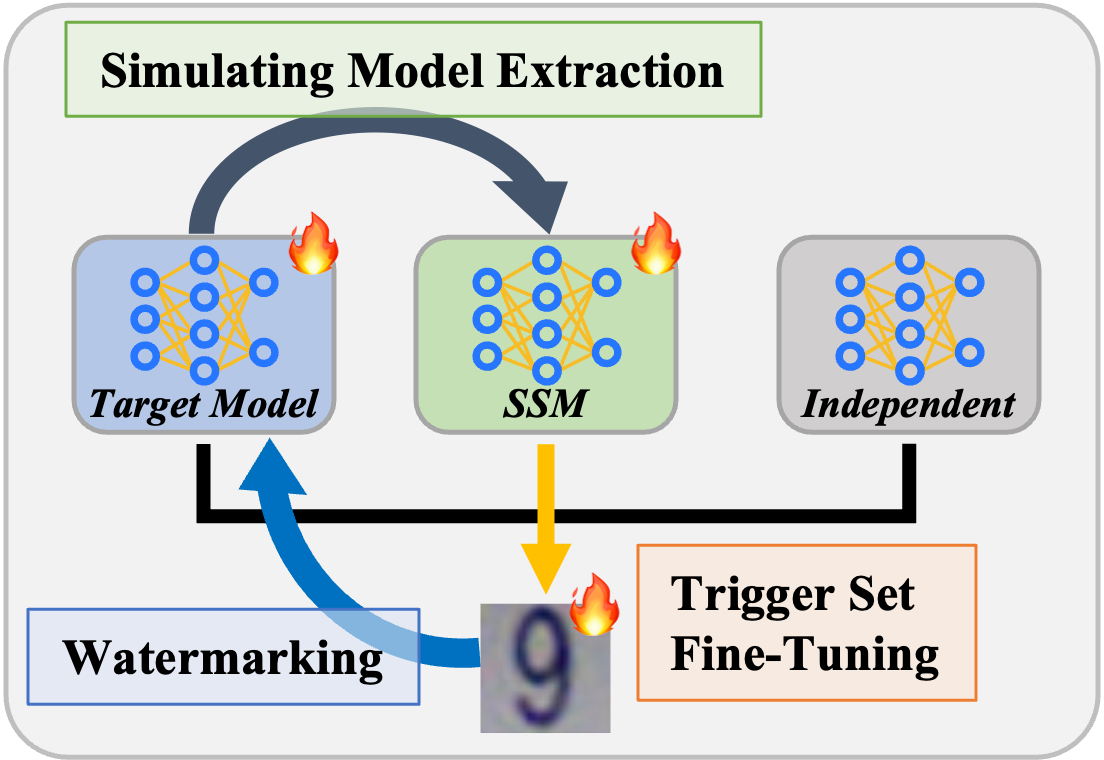}
        \label{subfig:SSW}
    }
	\hfill
	\subfloat[\emph{T2S} (Proposed)]{
        \includegraphics[width=0.36\linewidth,height=0.33\linewidth]{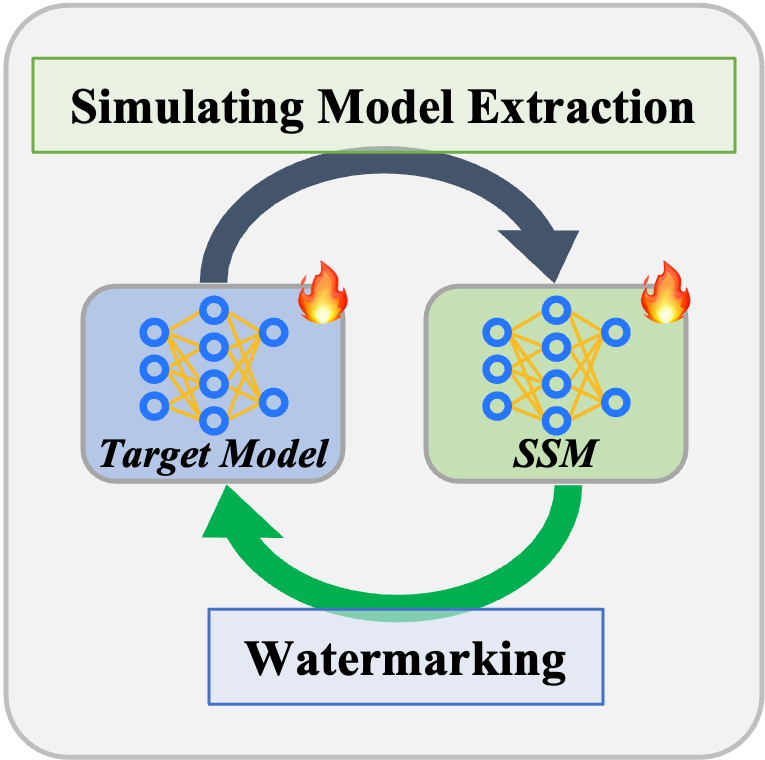}   \label{subfig:T2S}
    }
	\caption{Simulated Stolen Model (SSM) assisted watermarking. (a)  \emph{SSW} uses the SSM to fine-tune the trigger set (\textcolor{orange}{orange} arrow), which are then used to update the target model (\textcolor{blue}{blue} arrow); (b)  \emph{T2S} directly fine-tunes the target model based on feedback of SSM (\textcolor{green}{green} arrow).}
	\label{fig:illust_com} 
\end{figure}

Adversary simulation has been widely adopted in both attack and defense techniques \cite{ge21, zhou21}. In particular, the Symmetric Shadow
model based Watermarking (SSW) optimizes the trigger set for model watermarking by simulating the model extraction process \cite{tan23}.
As illustrated in Fig.\ref{subfig:SSW}, each round of \emph{SSW} consists of trigger set optimization, watermark embedding, and stealing simulation. First, the trigger set are optimized based on the watermarking loss of a simulated stolen model (SSM). The  target model is then watermarked with the updated trigger set and then distilled into the SSM to mimic model extraction. However, fine-tuning the trigger set to enhance watermark robustness may inadvertently compromise watermark uniqueness. To mitigate this issue, one or more independently pre-trained watermark-free surrogate models are employed as negative models to ensure low prediction accuracy for trigger sets, thereby preserving watermark distinctiveness.

\begin{figure*}[ht]
\vskip 0.2in
\begin{center}
\centerline{\includegraphics[width=\linewidth]{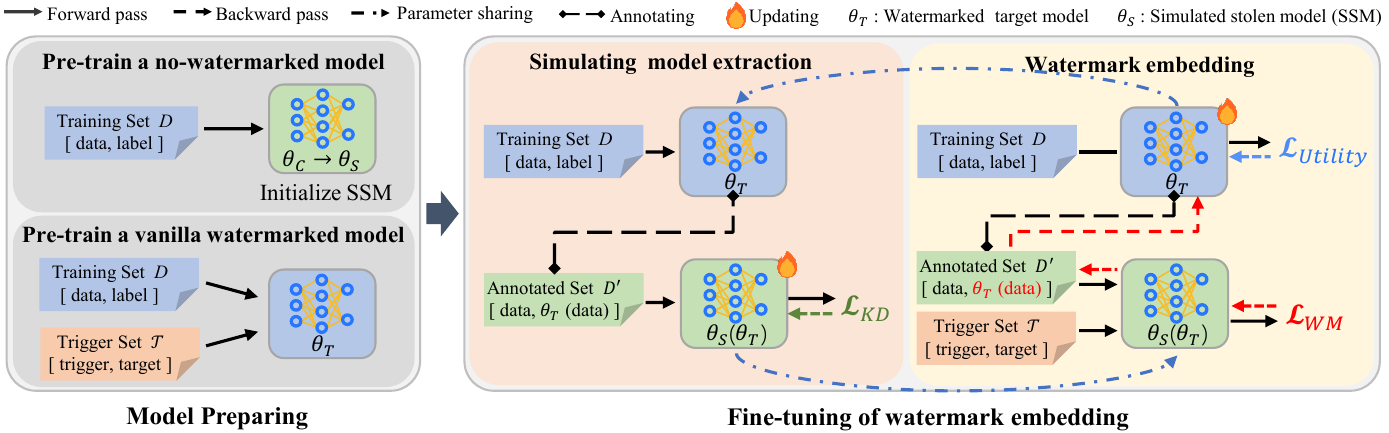}}
\caption{\textbf{Pipeline of the proposed \textit{T2S}.} The framework consists of a watermarked target model $\theta_T$ and a simulated stolen model (SSM) $\theta_S$. During the rehearsal process, $\theta_S$ imitates the model extraction behavior to generate feedback signals, while $\theta_T$ is fine-tuned using the second-order derivatives of the watermark loss $\mathcal{L}_{WM}$ with respect to its parameters (illustrated by the \textcolor{red}{red} dashed arrow). This meta-optimization enables $\theta_T$ to strengthen watermark robustness against potential extraction attacks.}
\label{fig:structure_T2S}
\end{center}
\vskip -0.2in
\end{figure*}

Although incorporating rehearsal of model extraction is effective, \emph{SSW} relies on an evolving trigger set to relay the current watermark transferability, as measured by the prediction quality of the trigger set in the SSM, back to the target model.
This raises a natural question: 

\textit{Can we establish a direct pathway for propagating feedback from the SSM to the target model?}

In our designed framework, the target model is updated directly using the loss computed from the current SSM on a fixed trigger set, eliminating the need for trigger set fine-tuning. This results in a simplified optimization cycle involving alternating updates between the target model and the SSM, as illustrated in Fig.\ref{subfig:T2S}. It is worth noting that that no independently trained models are required in this new framework when using a fixed trigger set.

\subsection{Tuning to Survive (\emph{T2S})}
\subsubsection{Overall structure}
Building on the above intuition, we propose a new model watermarking framework called \textbf{Tuning to Survive (\emph{T2S})}. The overall pipeline of this new approach, as shown in Fig.\ref{fig:structure_T2S}, consists of two stages: a pre-training phase and a fine-tuning phase.
After pre-training, the target model and SSM are initialized as a vanilla watermarked model and a watermark-free model, respectively. Both models are then fine-tuned alternately under our rehearsal-based joint optimization, where the key innovation is updating the target model with the \textbf{second-order derivatives} of the watermarking loss.

\begin{figure}[!t]
\centering
\includegraphics[width=0.85\linewidth]{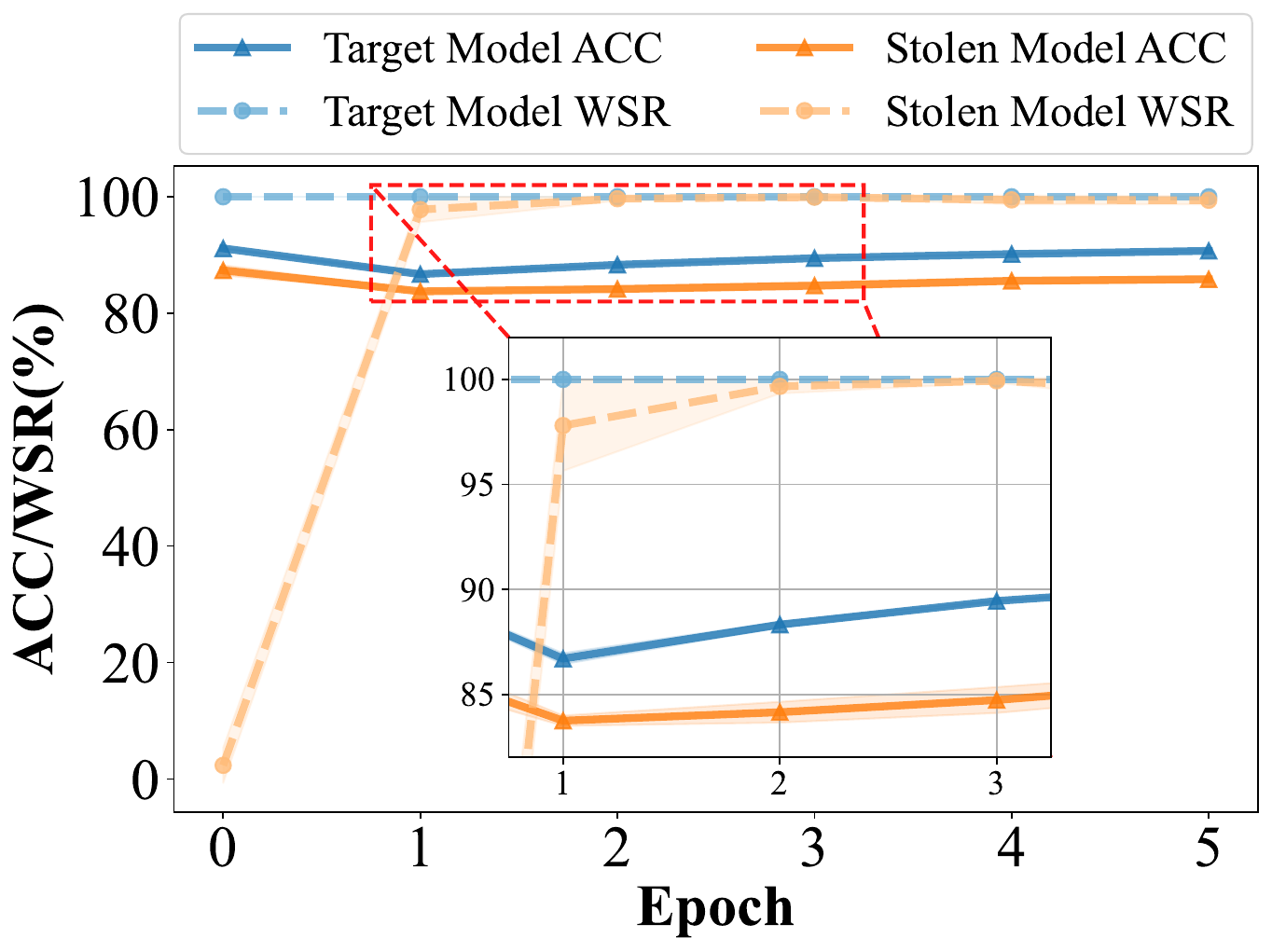}
\caption{ACC and WSR of target and stolen models over fine-tuning epochs with \textit{T2S} on CIFAR-10. The \textit{Knockoff} extraction attack is performed upon each epoch's checkpoint of the target model to train stolen models. }
\label{fig:epoch} 
\end{figure}

While pre-training successfully implants the watermark into the target model, it does not guarantee watermark transferability to stolen models. The fine-tuning phase reshapes the internal representation of watermark knowledge within the target model, enhancing its transferability during real model extraction attacks.
As shown in Fig.\ref{fig:epoch}, the watermark success rate (WSR) of the actual stolen model exhibits a sharp increase after just one epoch of fine-tuning, demonstrating that the direct feedback from the simulated stolen model is highly effective in enhancing resistance to model extraction.
While the clean-data accuracy (ACC) experiences a slight decline after the first epoch, it gradually recovers to its original level in subsequent epochs.

\subsubsection{Detailed formulation}
Assume that the defender has a labeled training set $(\mathbf{x},y)\in\mathcal{D}$ for classification learning and a trigger set $(\mathbf{x}_w,y_w)\in \mathcal{T}$ for watermarking. The goal of \emph{T2S} is to generate a target model $\theta_T$ implanted with extractable watermark assisted by jointing learning of $\theta_S$, a simulated stolen model (SSM). Notably, the proposed \emph{T2S} framework aims to provide a generic transferability enhancement solution, without being restricted to a specific type of trigger set. Next, we provide the detailed formulations of the key components.

\textbf{Model preparation.}
Before the rehearsal process, we first prepare a watermark-free target model $\theta_C$ that only learns the primary classification task, and use it to initialize the SSM, i.e., $\theta_S \leftarrow \theta_C$. Meanwhile, we pre-train a vanilla watermarked model $\theta_T$ by jointly learning the classification and watermarking tasks in a classic manner, i.e., using the cross-entropy loss over both normal and trigger samples:
\begin{align} \label{eq:mix}
\mathcal{L}_{mix}=\frac{1}{|B|}\sum_{(\mathbf{x},y)\in B}\mathcal{L}_{CE}(f(x;\theta_T),y),
\end{align}
where $B \in \mathcal{D} \cup \mathcal{T}$ denotes a mini-batch combining training and trigger samples.


\textbf{Simulation of model extraction.}
Given the current target model $\theta_T$, we train the SSM, i.e., $\theta_{S}$, by aligning its output with that of the target model on each training sample.
Specifically, for a coming mini-batch of training samples $B_D \in \mathcal{D}$ and a learning rate $\lambda$, the SSM is updated during fine-tuning with:
\begin{equation}
 \mathcal{L}_{KD}(\mathbf{x};\theta_T,\theta_S)=\frac{1}{|B_D|} \sum_{\mathbf{x}\in B_D}D_{KL}(f(\mathbf{x};\theta_T)||f(\mathbf{x};\theta_S))
\end{equation}

\begin{equation} \label{eq:S_update}
\theta_{S}\gets\theta_{S}-\lambda\frac{\partial\mathcal{L}_{KD}(\mathbf{x};\theta_T,\theta_S)}{\partial\theta_S}
\end{equation}
Here $\mathcal{L}_{KD}$ is the averaged KL divergence between the predictions of $\theta_S$ and $\theta_T$ for each input in the batch, simulating the knowledge distillation from the target model to the stolen model. In simulating model stealing, we adopted the same architecture for the SSM as the target model and leveraged the training data for model extraction. This setup mimics the worst-case scenario of real model stealing attacks, wherein the adversary possesses knowledge of the target model's architecture and access to the training data.

\textbf{Watermark fine-tuning.}
To quantify the transferability of the watermark currently embedded in the target model, we define the watermark loss $\mathcal{L}_{WM}$ of the simulated stolen model $\theta_S$ on the watermark trigger batch $B_\mathcal{T}$ as:
\begin{equation} \label{eq:WM}
\mathcal{L}_{WM}=\frac{1}{\mid B_\mathcal{T}\mid}\sum_{(\mathbf{x}_w,y_w)\in B_\mathcal{T}}\ \mathcal{L}_{CE}(f(x;\theta_S(\theta_T)),y_w).
\end{equation}
Here $\theta_S$ is a function of $\theta_T$ according to Eq.(\ref{eq:S_update}).
Taking the derivative of $\mathcal{L}_{WM}$ with respect to $\theta_T$, we have:
\begin{equation}
\frac{\partial\mathcal{L}_{WM}}{\partial\theta_{T}}=\frac{1}{\mid B_\mathcal{T}\mid}\sum_{(\mathbf{x}_w,y_w)\in B_\mathcal{T}}\ \frac{\partial\mathcal{L}_{CE}(f(\mathbf{x}_w;\theta_S(\theta_T)),y)}{\partial\theta_{T}}
\end{equation}
By applying the chain rule, the above partial derivative can be further expanded as:
\begin{equation} \label{eq:chain}
\frac{\partial\mathcal{L}_{WM}}{\partial\theta_{T}}=\frac{\partial\mathcal{L}_{WM}}{\partial\theta_{S}}\cdot\frac{\partial\theta_{S}}{\partial\theta_{T}}.
\end{equation}
The above formulation tells that we can update the target model $\theta_T$ with the watermarking loss calculated based on the simulated stolen model $\theta_S$. 
To ensure utility, meaning that the target model maintains its performance on the classification task, the
primary task loss remains included.
Specifically, for normal training data $B_D$, we have:
\begin{equation} \label{eq:utility}
\mathcal{L}_{Utility}= \frac{1}{\vert B_D \vert}\sum_{(\mathbf{x},y)\in B_D}\mathcal{L}_{CE}(f(\mathbf{x};\theta_T),y),
\end{equation}
and its gradient with respect to $\theta_T$:
\begin{equation}
\frac{\partial\mathcal{L}_{Utility}}{\partial\theta_{T}}=\frac{1}{\mid B_D\mid}\sum_{(x,y)\in B_D}\ \frac{\partial\mathcal{L}_{CE}(f(\mathbf{x};\theta_T),y)}{\partial\theta_{T}}\label{eq:grad_utility}
\end{equation}

Combing both the watermarking and utility losses, we have the final update rule for $\theta_T$ as below
\begin{equation}
\theta_{T}\gets\theta_{T}-\mu\left(\frac{\partial\mathcal{L}_{WM}}{\partial\theta_{T}}+\alpha\frac{\partial\mathcal{L}_{Utility}}{\partial\theta_{T}}\right) \label{eq:final}
\end{equation}
with $\frac{\partial\mathcal{L}_{WM}}{\partial\theta_{T}}$ and $\frac{\partial\mathcal{L}_{Utility}}{\partial\theta_{T}}$ being calculated by Eq.(\ref{eq:chain}) and Eq.(\ref{eq:grad_utility}), respectively.
Here, $\mu$ is the learning rate, and $\alpha$ is the coefficient that balances the importance of the classification performance compared to watermarking. Increasing $\alpha$ typically enhances the preservation of the primary classification task's performance, at the cost of reduced watermark resistance against model stealing (See Fig. \ref{fig:alpha_cifar10} and Fig.  \ref{fig:alpha_cifar100} for detailed experimental results with respect to $\alpha$). 
\subsection{Implementation details \label{alg}} 

\textbf{Fine-tuning vs. Training from scratch.}
Algorithm \ref{alg:T2S} provides the main steps of the proposed \textit{T2S} model watermarking approach. 
The default way to implement \emph{T2S} is to perform the rehearsal-based alternating fine-tuning after pre-training the target model on the mixed trigger and training data to be a vanilla watermarked model and the SSM on the normal training data as a watermark-free classification model. An alternative way is to pre-train both models as watermark-free ones. Compared to the default one, this strategy results in a slightly lower WSR for the target model. This implies that initial watermark knowledge contributes to enhanced robustness against model stealing after fine-tuning. Furthermore, this configuration ensures that the proposed approach can be directly applied in scenarios where a vanilla watermarked model is already available.

\begin{figure*}[!t]
    \centering
       \begin{minipage}{0.31\linewidth}
        \centering
        \includegraphics[width=\linewidth]{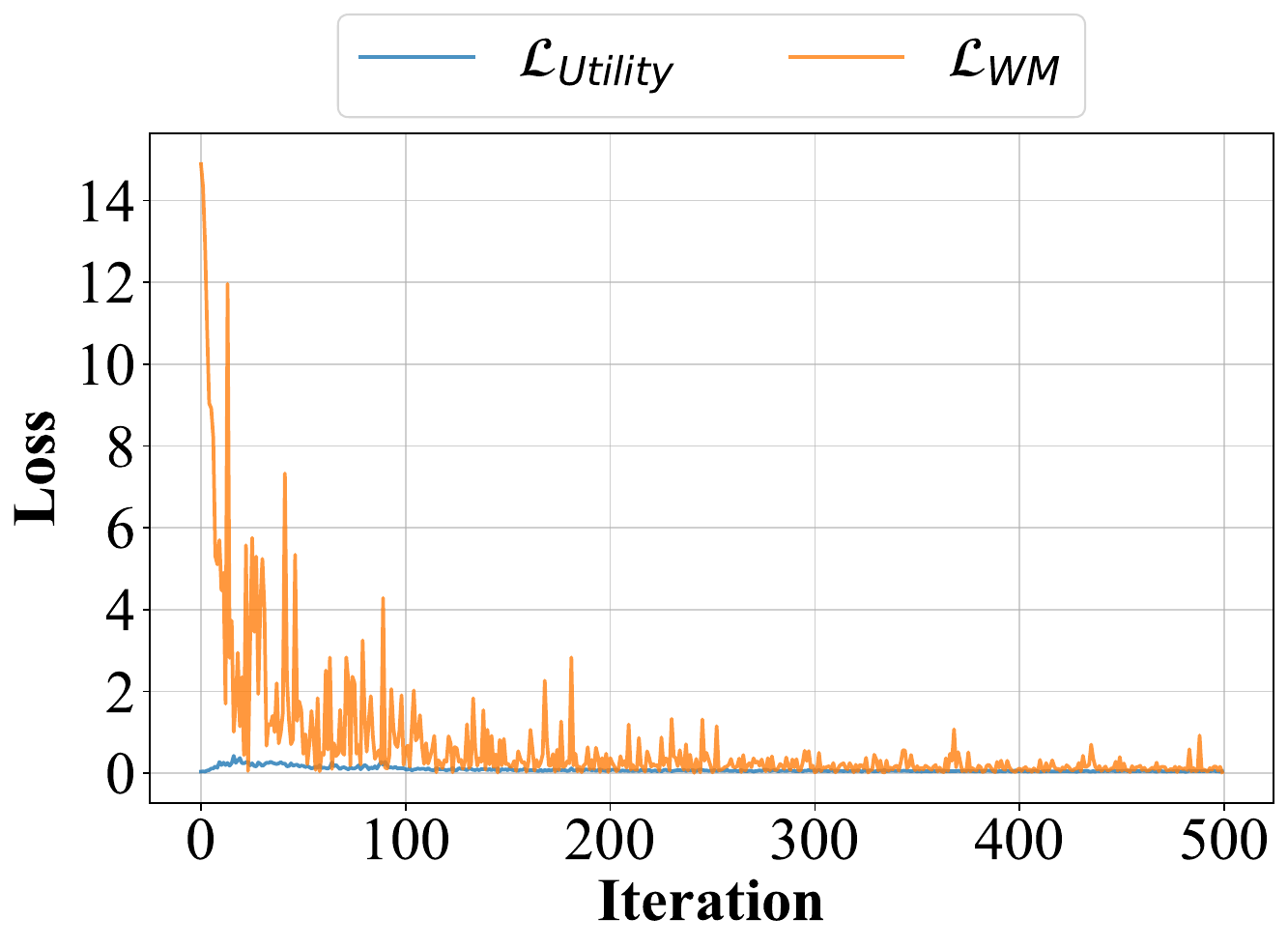}
        \subfloat{}
        \label{subfig:fine-tuning}
    \end{minipage}
    \hfill
    \begin{minipage}{0.31\linewidth}
        \centering
        \includegraphics[width=\linewidth]{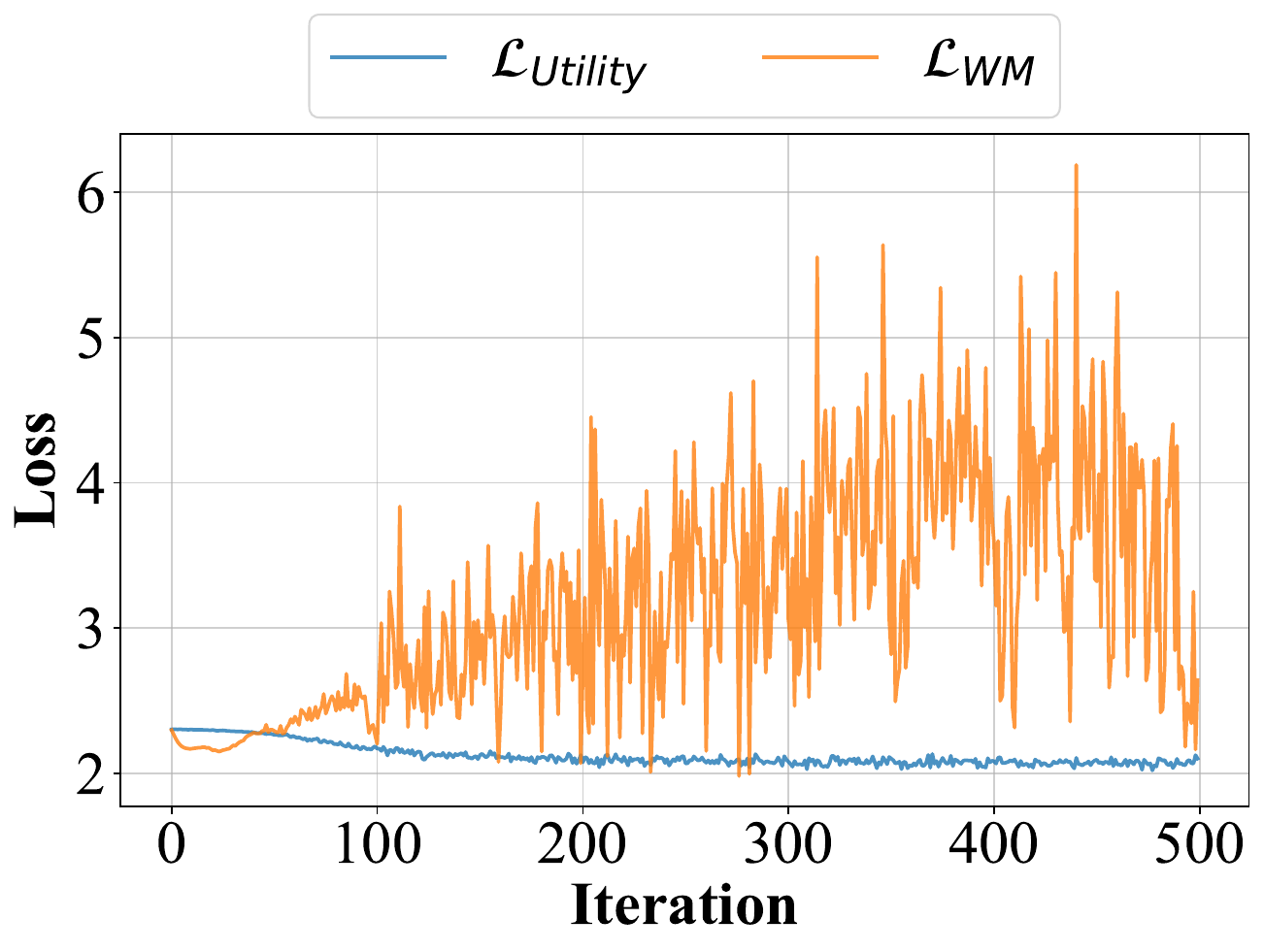}
        \subfloat{}
        \label{subfig:}
    \end{minipage}
   \hfill
    \begin{minipage}{0.31\linewidth}
        \centering
        \includegraphics[width=\linewidth]{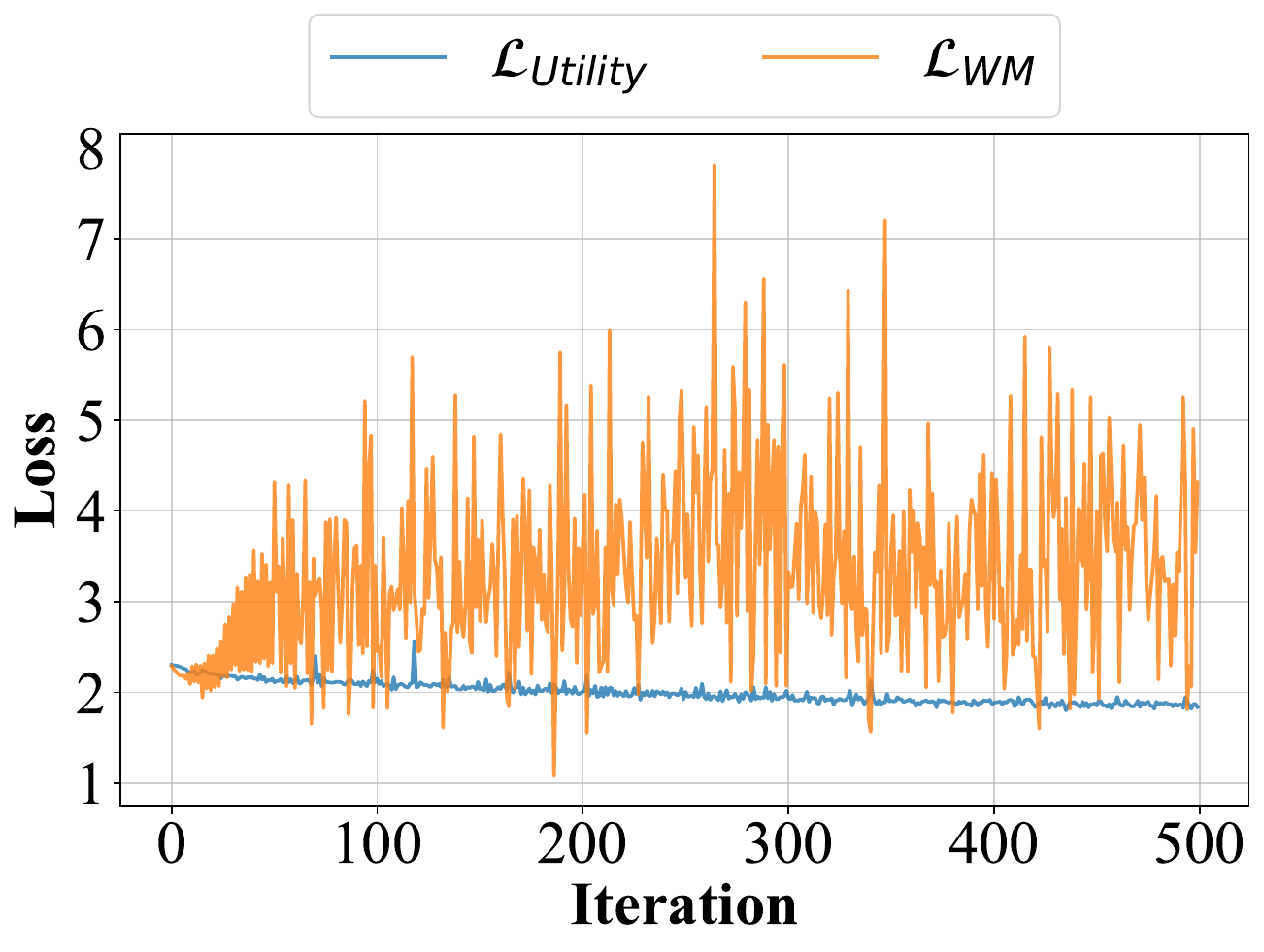}
        \subfloat{}
        \label{subfig:scratch}
    \end{minipage}
    \caption{Comparison of learning curves between fine-tuning and training from scratch with different learning rates. \textbf{left:} pre-training and fine-tuning; \textbf{middle:} training from scratch with learning rate 0.01 \textbf{right:} training from scratch with learning rate 0.1. }
    \label{fig:learning_curves} 
\end{figure*}

We also tried training from scratch, i.e., both the target model and SSM randomly initialized. Convergence issues occur in this setting, as seen from learning curves in Fig. \ref{fig:learning_curves} (the middle and right subplots). Specifically, for CIFAR-10, the model plateaus at 36.47\% accuracy (ACC) and 23.00\% weighted success rate (WSR) after 30 training epochs, a duration equivalent to completing both pre-training and fine-tuning in our proposed framework. 

Pre-training is thus necessary to ensure the stability and effectiveness of subsequent joint watermarking and extraction simulation in \textit{T2S}. 
This phased implementation confines the computationally intensive simulation to the fine-tuning phase (completing within 1-2 epochs), largely reducing total overhead compared to end-to-end training.  It also enables us to apply potential advanced fine-tuning techniques with better computation and data efficiency, e.g., using a small subset for fine-tuning, enhancing the scalability of our approach to large datasets.

\textbf{Watermark trigger set.}
Since the proposed \textit{T2S} framework is not formulated with a specific type of trigger set, it is applicable to 
various trigger sets designed for backdoor-based model watermarking. In our experiments, we apply our watermark tuning framework with three different types of trigger sets used in the literature, namely OOD \cite{jia21}, Mix  \cite{lv24}, and feature-based \cite{zeng23}. OOD uses out-of-distribution images as trigger inputs and labels them to a target class, while Mix combines training images from two different source classes of the training data to synthesize trigger samples. 
The feature-based trigger samples are generated by stamping a class-related trigger into samples of a selected source class. Specifically, given a pretrained watermark-free model $f_{\theta_C}$, a feature-based trigger $\delta$ of the same size as the input image is optimized using samples $D_s$ from the source class $s$ to minimize the following loss
\[
\delta^* \leftarrow \min \sum_{(x,y)\in D_s}  \mathcal{L}(f_{\theta_C}(x+\delta),y)
\]
The learned trigger $\delta^*$ captures class-specific features of the source class. 
 


\begin{algorithm}[!t]
    \caption{Watermarking by Rehearsal (\textit{T2S})}
    \label{alg:T2S}
    \renewcommand{\algorithmicrequire}{\textbf{Input:}}
    \renewcommand{\algorithmicensure}{\textbf{Output:}}

    \begin{algorithmic}[1]
        \Require Training set $D$, trigger set $\mathcal{T}$, coefficient $\alpha$
        \Ensure Watermarked model $\theta_T$
        
        \Statex \textbf{Model Preparation:}
        \State Train a watermark-free target model $\theta_C$ on $D$
        \State Initialize simulated stolen model (SSM): $\theta_S \leftarrow \theta_C$
        \State Pre-train a vanilla watermarked target model $\theta_T$ on $D \cup \mathcal{T}$ with Eq.~(\ref{eq:mix})

        \While{not converged or still have training budget}
            \Statex \textbf{(1) Simulate model stealing to update $\theta_S$:}
            \State Sample a batch $B_D \subset D$ 
            \State For each $\mathbf{x} \in B_D$, obtain pseudo-label $y' = f(\mathbf{x}; \theta_T)$
             \State Compute the distillation loss $\mathcal{L}_{KD}$ and update $\theta_S$ according to Eq.~(\ref{eq:S_update})
            \State Compute the utility loss $\mathcal{L}_{Utility}$ with Eq.~(\ref{eq:utility})
            
            \Statex \textbf{(2) Watermark fine-tuning in $\theta_T$:}
            \State Sample a batch $B_{\mathcal{T}} \subset \mathcal{T}$
            \State Compute watermark loss $\mathcal{L}_{WM}$ by Eq.~(\ref{eq:WM})
            \State Update $\theta_T$ based on the two losses using Eq.~(\ref{eq:final})
        \EndWhile
    \end{algorithmic}
\end{algorithm}

\section{Experiment}
\label{sec:experiment}
In this section, we present a comprehensive evaluation of the proposed \emph{T2S} framework for watermarking image classification models. We first provide the details of experimental settings, including the model extraction attacks considered, the baseline watermarking approaches for comparison, as well as training details and hyperparameter settings. We then focus on the examining the robustness of \emph{T2S} from the following three aspects:
\begin{itemize}
    \item Robustness against model extraction. We evaluate the impact of model watermarking to the target model's classification performance, and watermark extractibility‌ in stolen models under different model stealing attacks, sensitivity to variations in model architectures, as well as dependence on the choice of query datasets.
    \item Further watermark removal after stealing. We then investigate the resilience of the \emph{T2S} watermark transferred to the stolen model under potential following removal strategies, such as secondary extraction, quantization, and pruning. 
    \item Robust to compression of target model. Additionally, we assess the robustness of \emph{T2S} when the watermarked target model undergoes compression before deployment.
\end{itemize}
 Ablation studies are also conducted to evaluate the impact of key factors, including the weighting parameter $\alpha$, and the selection of source and target classes. Finally, we provide additional analyses on practical aspects such as computational time and storage requirements. 






\subsection{Experiment Settings} \label{App:exp_setting}

\textbf{Datasets, target-model architectures, and evaluation.}
Our evaluations are mainly carried out with image classification tasks on two standard image datasets, CIFAR-10 and CIFAR-100 \cite{cifar10}. We use ResNet-18 \cite{resnet} as the target model architecture for CIFAR-10 and CIFAR-100. Model watermarking performance is evaluated using test accuracy (ACC) and watermark success rate (WSR), with all results reported as means and standard deviations over \textit{five} random trials. 


\textbf{Model stealing attacks.\label{extraction_attack_des}}
We evaluate the robustness of model watermarking approaches against representative black-box extraction-based model stealing attacks, namely \emph{Knockoff} \cite{orekondy19knock} (both soft label and hard label) and \emph{DFME} \cite{truong21}. \textit{Knockoff} leverages an unlabeled proxy dataset to query the target model and trains the stolen model to match the labels returned by the target model. We consider both soft-label and hard-label settings, corresponding to scenarios where the target model returns either a probability distribution over all classes or a single class for each query. The proxy dataset can be either in-domain or out-of-domain, depending on the resources available to the attacker.  
\textit{DFME} is a data-free model extraction approach that jointly trains a generative model to synthesize queries along with stolen model learning. The performance of the stolen model is highly dependent on the quality and quantity of the queries. For \textit{Knockoff}, we use 50K queries for CIFAR-10 and CIFAR-100, following prior work~\cite{jia21}. Following the original setting of \textit{DFME}, 20M synthetic queries are used.

In most experiments, we use the same network architecture as the target model for stolen models, and use in-distribution images sampled from the target model's training set as query data for \emph{Knockoff}. We investigate the impact of both the query dataset and stolen model architecture in separate experiments.
 
\textbf{Compared approaches.} We compare \emph{T2S} with other three representative model-watermarking methods, including \textit{Content} \cite{zhang18}, \textit{EWE} \cite{jia21}, and two variants of \textit{SSW}\cite{tan23}, where SSW-P indicates that the target and stolen models are fine-tuned based on pre-trained models, and SSW-S means trained from scratch. 
For \textit{SSW}\cite{tan23}, we conducted experiments using author published source code. For \textit{Content} \cite{zhang18} and \textit{EWE} \cite{jia21}, we conducted experiments using the source code provided by the survey paper \cite{lukas22}. 
We also attempted to compare our results with another approach proposed in \cite{lv24}. However, we encountered significant challenges in establishing a reliable baseline for this approach \footnote{The original paper's reported results exhibit substantial variance under identical experimental conditions, e.g., 84.40\% WSR in Table 1 versus 71.05\% in Table 4 for CIFAR-10 \cite{lv24}. Our reproduction using the authors' released code gives even lower WSR than their most reported values.}.

\textbf{Training details and hyperparameter settings.}
During the model preparing stage, i.e., pre-training both the watermark-free and vanilla watermarked target models, we use the SGD optimizer with cosine annealing learning rate scheduling. The initial learning rate is 0.1, momentum is 0.9, and weight decay is $5\times10^{-4}$. Training is performed for 40 epochs with a batch size of 256.
During the rehearsal process, the simulated stolen model is updated in an autograd manner with a learning rate of $1\times10^{-3}$, while the target model is fine-tuned using the SGD optimizer with an initial learning rate of $5\times10^{-3}$ for 3 epochs. The batch sizes are set to $B_D=B_\mathcal{T}=100$.

By default, \textit{T2S} employs a feature-based trigger set. For CIFAR-10, we select ``truck'' and ``frog'' as the source and target classes, respectively, and randomly sample 500 host images to which the learned source-class trigger is attached. The CIFAR-100 trigger set is constructed in a similar manner, containing 500 samples with ``apple'' as the source class and ``willow tree'' as the target class.  See Section \ref{sec:triggertype} for more results across various source–target class combinations. The weighting coefficient $\alpha$ is by default set to 50 and 15 for CIFAR-10 and CIFAR-100, respectively. 
We study the impact of $\alpha$ in Section \ref{sec:alpha} (See Fig. \ref{fig:alpha_cifar10} and \ref{fig:alpha_cifar100} for results of different values of $\alpha$).





\subsection{Robustness against model extraction}

\begin{table}
    \centering
    \caption{\textbf{Impact on target model after watermarking.}}
    \label{tab:WMtargetModel}
   \resizebox{\linewidth}{!}{
    \begin{tabular}{cccccc}
			\toprule
			\multirow{2}{*}{Task} & \multirow{2}{*}{Method} & \multicolumn{2}{c}{No-watermark} & \multicolumn{2}{c}{Watermarked} \\
			\cmidrule{3-6}
			& & ACC (\%) & WSR (\%) & ACC (\%) & WSR (\%) \\
			\midrule
			\multirow{5}{*}{CIFAR-10}
            & Content & \multirow{5}{*}{90.05±0.15} 
                    & 0.10±0.22 & 89.32±0.46
& 100±0.00
\\
			& EWE &  & 0.00±0.00 & 89.15±0.34
& 100±0.00
\\
			& SSW-P &  & 3.00±2.45 & 89.52±0.12
& 100±0.00
\\
            & SSW-S &  & 0.60±0.89 & 89.01±0.27
& 100±0.00
\\
			& T2S &  & 0.32±0.50 & 89.54±0.25
& 100±0.00
\\
			\midrule
			\multirow{5}{*}{CIFAR-100} 
            & Content & \multirow{5}{*}{70.28±0.21}& 0.00±0.00
& 70.31±0.16
& 100±0.00
\\
			& EWE &  & 0.00±0.00
& 70.12±0.38
& 100±0.00
\\
			& SSW-P &  & 0.26±0.62
& 70.52±0.51
& 100±0.00
\\
            & SSW-S &  & 0.22±0.47
& 70.87±0.25
& 100±0.00
\\
			& T2S &  & 0.12±0.33
& 70.02±0.37
& 100±0.00
\\
            \midrule
			\multirow{5}{*}{Tiny-ImageNet} 
            & Content & \multirow{5}{*}{50.15±0.34}& 0.00±0.00
& 49.76±0.74
& 100±0.00
\\
			& EWE &  & 0.33±0.57
& 49.92±0.25
& 100±0.00
\\
			& SSW-P &  & 0.57±0.24
& 50.63±0.13
& 100±0.00
\\
            & SSW-S &  & 0.22±0.32
& 50.34±0.31
& 100±0.00
\\
			& T2S &  & 0.20±0.20
& 50.01±0.10
& 100±0.00
\\
			\bottomrule
		\end{tabular}    
   } 
    \vskip -0.1in
\end{table}

\begin{table*}
\centering
\caption{\textbf{Robustness of model watermarking against model extraction.} Performance of stolen models (ACC, WSR, \%) extracted (with three attacks) from target models (watermarked by five methods).}\label{tab:model extraction}
\begin{tabular}{ccccccc}
\toprule
			  \multirow{2}{*}{Method} & \multicolumn{2}{c}{Knockoff (soft label)} & \multicolumn{2}{c}{Knockoff (hard label)} & \multicolumn{2}{c}{DFME} \\
			\cmidrule{2-7}
			 & ACC  & WSR  & ACC  & WSR  & ACC & WSR  \\
			\midrule
	    \textbf{CIFAR-10}\\\midrule
             Content\cite{zhang18}&   85.94±0.13  &      1.20±0.89       &   81.25±0.29  &   0.00±0.00   &   89.32±0.61  &   84.80±6.14  \\
		EWE\cite{jia21}    &   87.10±0.38  &       15.00±9.12      &   83.46±0.46  &   0.60±0.80   &   88.05±0.48  &   87.40±7.31  \\
		SSW-P\cite{tan23}  &   85.02±0.15  &       65.25±16.19      &   83.13±1.47  &   21.60±11.61   &   85.39±0.83  &   \textbf{100.00±0.00}  \\
        SSW-S\cite{tan23}  &   85.34±0.14  &       89.60±6.91      &   83.19±1.73  &   72.40±6.35   &   86.18±0.71  &   \textbf{100.00±0.00}  \\
		T2S  &   86.87±0.55  &\textbf{99.87±0.23}    &   82.01±0.70  & \textbf{97.56±2.97}  &  85.27±0.76  &\textbf{100.00±0.00}  \\
			\midrule
			\textbf{CIFAR-100}\\\midrule
             Content\cite{zhang18}&    65.07±0.33  &   34.69±9.90   &   58.76±0.65  &   0.00±0.00   &   33.64±0.70  &   72.03±10.28  \\
		EWE\cite{jia21}    &    64.93±0.95 &	50.76±16.91&	57.87±1.76&	3.62±2.01&	35.51±0.76&	75.44±11.53  \\
		SSW-P\cite{tan23}  &   64.98±0.65&	91.24±3.71&	58.98±0.26&	15.41±6.96&	36.83±0.93&	\textbf{100.00±0.00} \\
        SSW-S\cite{tan23}  &   65.15±0.37&	92.73±3.15&	59.71±1.28&	34.75±10.43&	35.80±1.31&	\textbf{100.00±0.00}  \\
		T2S  &   65.27±0.45&	\textbf{95.42±0.47}&	59.49±0.51&	\textbf{68.68±3.61}&	30.82±4.04&	\textbf{100.00±0.00}  \\
        \midrule
			\textbf{Tiny-ImageNet}\\\midrule
             Content\cite{zhang18}&    47.98±0.13  &   1.07±1.85    &   36.13±0.09  &   0.00±0.00   &   11.78±3.84  &   65.11±15.81  \\
		EWE\cite{jia21}           &    48.96±0.44  &  43.76±20.55   &	36.31±0.10  &	10.37±13.82   &	11.66±0.39  &	69.73±23.56  \\
		SSW-P\cite{tan23}         &   48.54±0.09   &  84.83±7.82    &	36.85±0.52  &	86.71±8.24    &	11.76±2.73    &	93.50±6.85 \\
        SSW-S\cite{tan23}         &   48.57±0.59   &    89.28±10.11 &	37.01±0.14    &	88.02±8.97   &	12.35±2.12    &	98.72±9.56  \\
		T2S                       &   48.46±0.21   &	\textbf{93.93±1.42}   &  36.39±0.42    &	\textbf{90.40±4.11 } &	12.19±1.01    &	\textbf{99.20±0.80 } \\
\bottomrule
\end{tabular}
\end{table*}


\subsubsection{Watermarked target models}
We first evaluate the impact of watermarking on the primary-task performance of the target model. 
Table \ref{tab:WMtargetModel} shows that our watermarking approach results in negligible performance drops on the main task while achieving a 100\% success rate for triggering the watermark implanted in the target model. In fact, all the five approaches perform well in watermarking the target models, achieving small ACC decreases and high WSRs, indicating that watermarking a model itself is relatively easy.

\begin{table}[t]
    \centering
    \caption{\textbf{Impact of architecture difference.} ACC and WSR (\%) of \textit{Knockoff} (soft label) stolen models from \textit{T2S}-watermarked target models on CIFAR-10 and Tiny-ImageNet.}
    \label{tab:different model}
   \resizebox{1\linewidth}{!}{  
    \begin{tabular}{cccccc}
			\toprule
			 &\multicolumn{2}{c}{ResNet-18 (CIFAR-10)}&\multicolumn{2}{c}{MobileViT-s (Tiny-ImageNet)}\\\cmidrule(r){2-3}\cmidrule{4-5}\
             Stolen Model 
            & ACC & WSR & ACC & WSR\\
            \midrule
               ResNet-18     &   86.87±0.55   &   99.87±0.23             &   53.46±0.66   &   \textbf{96.60±1.64} \\
			    WRN-16-4     &   88.96±0.45   &   \textbf{99.96±0.08}       &   51.49±0.91   &   89.93±1.62\\
			    ViT-s        &   70.81±0.20   &  80.53±2.20   &   41.02±0.24   &   86.07±4.94\\
			    MobileViT-s  &   83.04±0.52   &   85.20±4.80             &   48.46±0.21   &   93.93±1.42\\
			\bottomrule
		\end{tabular}
   }
\end{table}

\subsubsection{Robustness against model stealing}

Table~\ref{tab:model extraction} compares the extractibility of the watermark in stolen models whose corresponding target models are watermarked using five different approaches. Since all methods achieve comparable test accuracies (ACC), we focus our analysis on watermark success rates (WSR). As seen from the table, it is relatively easy to preserve watermarks against the \textit{DFME} attack, as all watermarking methods achieve more than 70\% WSR in stolen models, with \textit{SSW} and \textit{T2S} reaching nearly 100\%. However, the first two approaches struggle to retain watermarks under the \textit{Knockoff} attack, especially in the hard-label setting, where WSR drops below 11\%. The use of extraction simulation substantially improves watermark retention against \textit{Knockoff}, as evidenced by the high WSRs achieved by the three rehearsal-based methods, namely \textit{SSW-S}, \textit{SSW-S} and \textit{T2S}. Notably, our \textit{T2S} maintains strong watermark robustness, achieving more than 97\% WSR on CIFAR-10 and 68\% on CIFAR-100, even under the challenging \textit{Knockoff} (hard-label) setting. In the following analytical experiments, we therefore focus on evaluating \textit{T2S} under the \textit{Knockoff} (soft-label) scenario.

%
\subsubsection{Sensitivity to model architectures}
In our initial experiments, we used stolen models with identical architecture to the target models. However, in realistic attack scenarios, the attacker typically lacks knowledge of the target model's architecture. We therefore evaluate our approach's performance using various architectures for the stolen models. 

Table \ref{tab:different model} gives the classification performance and watermark success rate of stolen models of four different architectures that are extracted from the CIFAR-10 model and Tiny-ImageNet model with \textit{Knockoff} (soft label).
Overall, the proposed watermarking method maintains reasonable model performance while exhibiting strong robustness and transferability against cross-architecture model extraction scenarios. This observation indicates that architectural differences do not directly affect $T2S$. 

For the three datasets used in our study, convolutional networks (ResNet-18 and WRN-16-4) are more favorable than transformer-based architectures (e.g., ViT-s and MobileViT-s), as  transformers often require larger, higher-quality datasets to fully leverage their capabilities. This explains the inferior performance of the last two rows when using transformer-based models.


\begin{table}[t]
    \centering
    \caption{\textbf{Impact of query datasets.} ACC and WSR (\%) of \textit{Knockoff} (soft label) stolen models from \textit{T2S}-watermarked ResNet-18 targets.}
    \label{tab:different dataset}
    \resizebox{1\linewidth}{!}{  
    \begin{tabular}{ccccc}
			\toprule
			&\multicolumn{2}{c}{CIFAR-10 (ResNet-18)}&\multicolumn{2}{c}{CIFAR-100 (ResNet-18)}\\\cmidrule(r){2-3}\cmidrule{4-5}
            Query data & ACC & WSR  & ACC & WSR \\
            \midrule
            CIFAR-10         &   86.87±0.55     &   99.87±0.23              &   59.24±0.73      &   94.48±4.65\\
			CIFAR-100        &   82.30±0.48      &   99.76±0.39              &   65.27±0.45      &   95.42±0.47\\
			STL-10           &   87.78±0.53      &   \textbf{100.00±0.00}    &   64.16±0.39      &   97.60±2.11\\
			TinyImageNet     &   87.49±0.63      &   \textbf{100.00±0.00}    &  65.44±0.14       &   \textbf{98.52±1.05}\\
			\bottomrule
		\end{tabular}
    } 
    \vskip -0.1in
\end{table}

\subsubsection{Sensitivity to query datasets} 
As shown in Table \ref{tab:different dataset}, the stolen models extracted using queries from all four datasets consistently achieve high WSRs, demonstrating the robustness of the \textit{T2S}-embedded watermark against variations in the query data distribution. 
Interestingly, using out-of-distribution (OOD) queries even results in higher WSRs than using in-distribution ones. A similar observation has been reported and analyzed by \cite{Zong24IPremover} in the context of backdoor removal. 
 One possible explanation is that the competing objectives during the target model’s training introduce an optimization trade-off — learning the primary classification task for the target distribution may partially offset the ability to encode the watermark, leading to a reduced WSR.
 Since the model has never been exposed to the OOD distribution, querying with OOD samples does not interfere with the embedded watermark knowledge, thereby producing higher WSRs.

\subsection{Watermark removal upon stolen model}
We now examine whether the $T2S$-embedded watermark transferred into the stolen model still persists when subjected to additional watermark removal attempts. 
To assess this, we apply three widely used watermark removal techniques, namely second extraction, quantization\cite{hubara18} and pruning \cite{zhu17} upon stolen models.  Second extraction means that the adversary user applies \textit{Knockoff} (soft label) to extract the stolen model into another new model. Weight quantization and pruning are widely used model compression techniques that reduce model size by employing lower numerical precision (e.g., 8-bit or 4-bit) or setting small weights to zero.



\begin{table}[t]
    \centering
    \caption{\textbf{Robustness to extraction-based watermark removal.} ACC and WSR (\%) after 2nd \textit{Knockoff} extraction with in/out-of-distribution queries. Mean changes are marked by $\uparrow$ and $\downarrow$.}
    \label{tab:stolen_extraction}
    \resizebox{\linewidth}{!}{  
    \begin{tabular}{ccc}
        \textbf{CIFAR-10}\\\toprule
        Query data & ACC & WSR  \\
        \midrule
        CIFAR-10 & 86.84±0.63 ($\downarrow$ 0.03) & 99.93±0.12 ($\uparrow$ 0.06)  \\
        STL-10 & 87.99±0.19 ($\downarrow$ 0.21)  & 100.00±0.00 ($\uparrow$ 0.00)       \\
        \bottomrule\\
         \textbf{CIFAR-100}\\\toprule
         Query data & ACC & WSR \\
        \midrule
        CIFAR-100 &  66.56±0.23 ($\uparrow$ 0.76) & 96.73±0.76 ($\downarrow$ 0.67)\\
        TinyImageNet & 65.89±0.57 ($\uparrow$ 0.45) & 100.00±0.00 ($\uparrow$ 0.00) \\
        
        \bottomrule
    \end{tabular}
    }
\end{table}

\begin{figure*}[!t]
    \centering
    \begin{minipage}{0.4\linewidth}
        \centering
        \includegraphics[width=\linewidth]{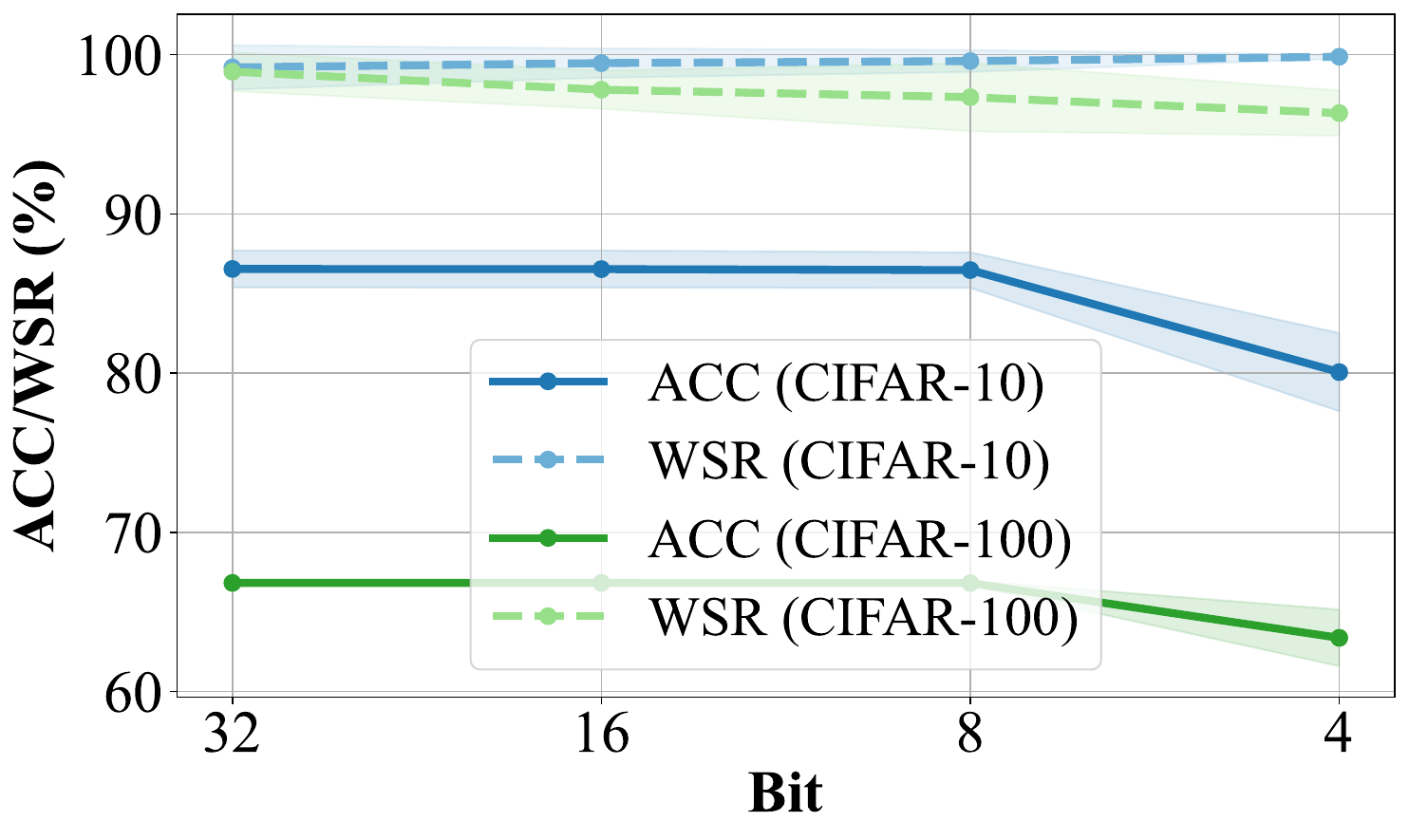}
        \subfloat{Weight Quantization}
        \label{subfig:stolen_quantization}
    \end{minipage}
    \hspace{0.1\linewidth} 
    \begin{minipage}{0.4\linewidth}
        \centering
        \includegraphics[width=\linewidth]{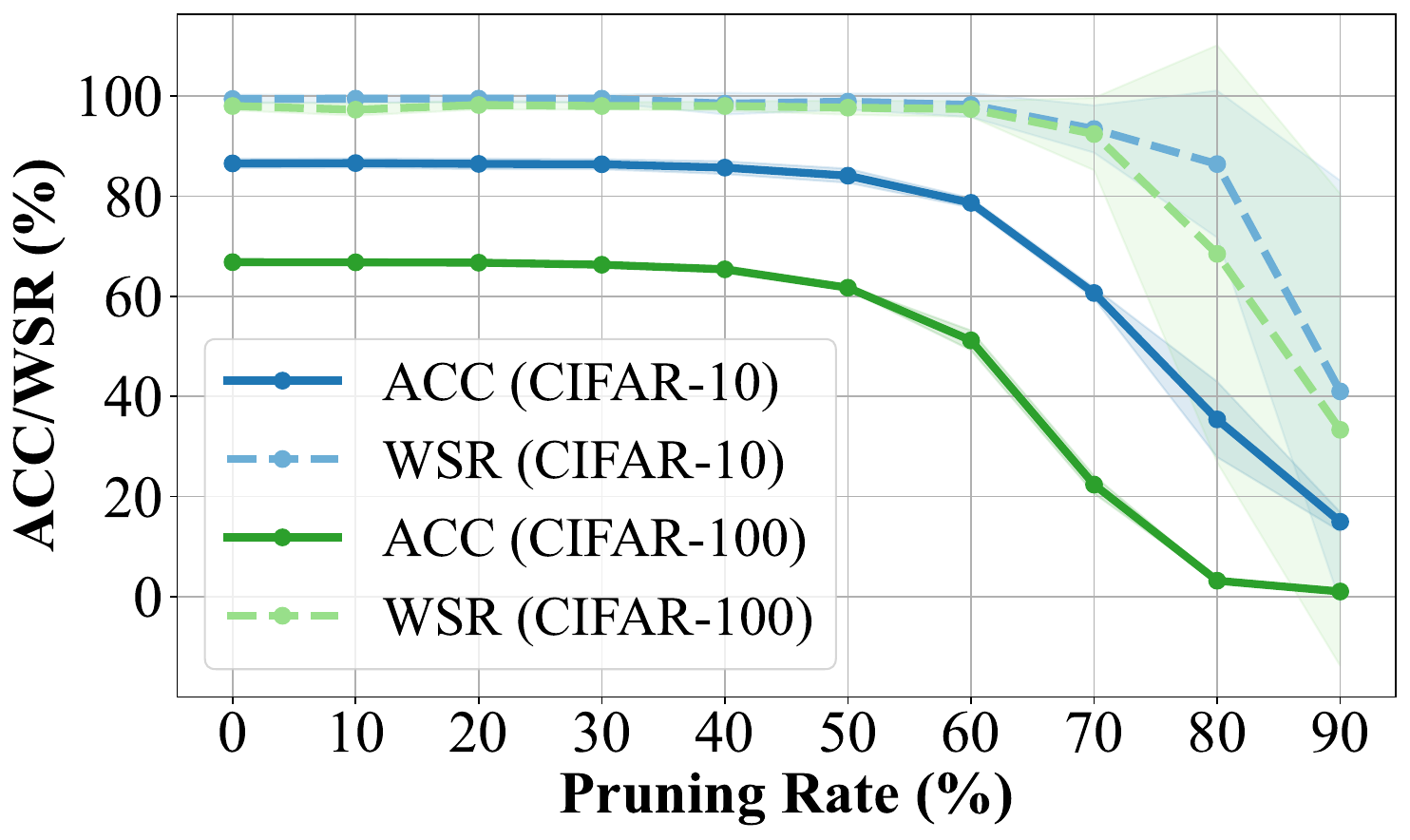}
        \subfloat{Pruning}
        \label{subfig:stolen_pruning}
    \end{minipage}
    \caption{Results of stolen model after model compression. (a) results with different levels of weight quantization. (b) results with varying pruning rates.}
    \label{fig:stolen_quantization_pruning}
\end{figure*}

\textbf{Second extraction.}
Table \ref{tab:stolen_extraction} gives the results of watermark resilience when the stolen model undergoes secondary model extraction. Decrease in both ACCs and WSRs keeps within 0.5\% in all cases, showing that further extraction is ineffective for removing the watermark in the stolen model.


\textbf{Quantization and pruning.}
Results in Fig.\ref{fig:stolen_quantization_pruning} demonstrate the high persistence of watermarks in stolen models against two representative model compression techniques, i.e., quantization and pruning. 
It is shown that both the classification performance and watermark success rate are nearly unaffected with more than 8 bit precision or a pruning rate $<$ 50\%.
 However, the classification accuracy begins to decline sharply when reducing the precision to 4-bit or  pruning more than 50\% weights. Nevertheless, the WSR remains above 95\% for both datasets even under 4-bit quantization, and remains above 90\% until the pruning rate surpasses 70\%. 
This indicates that the watermark remains largely intact under practically acceptable compression levels.



\subsection{Robustness against target model compression}
Model compression techniques are widely used to reduce the
size of trained models before deployment, optimizing their
computational efficiency and storage requirements while
preserving performance. However, these techniques may
compromise the embedded watermarks, causing difficulties in tracing the watermarked model. Therefore, it is important for the implanted watermarks to be robustness to model compression.
We investigate the impact of three popular model compression methods, including distillation, quantization and pruning on $T2S$ watermarked models.
\begin{table}[t]
    \centering
    \caption{\textbf{Performance of watermarked models after distillation.}}
    \label{tab:distill_compression}
    \begin{tabular}{cccc}
			\hline
			\multirow{2}{*}{\textbf{Dataset}} & \multirow{2}{*}{\textbf{Method}} & \multicolumn{2}{c}{\textbf{Distillation}}  \\
            \cline{3-4}
            & & \textbf{ACC (\%)} & \textbf{WSR (\%)} \\
            \hline
            \multirow{3}{*}{CIFAR-10}
            &   SSW-P       &   89.39±0.18  &   60.36±16.26 \\
			&   SSW-S        &   89.42±0.15   &   60.00±31.75 \\
			&   T2S        &   90.18±0.22   &   \textbf{73.87±18.85} \\
			\hline
            \multirow{3}{*}{CIFAR-100}
            &    SSW-P  &   67.40±0.29  &   56.60±21.77 \\
			&    SSW-S   &   68.01±0.23   &  45.40±18.26 \\
			&    T2S   &   68.13±0.38   &   \textbf{80.67±3.95} \\
			\hline
		\end{tabular}
\end{table}



\begin{figure*}[!t]
	\centering
    \begin{minipage}{0.4\linewidth}
        \centering
        \includegraphics[width=\linewidth]{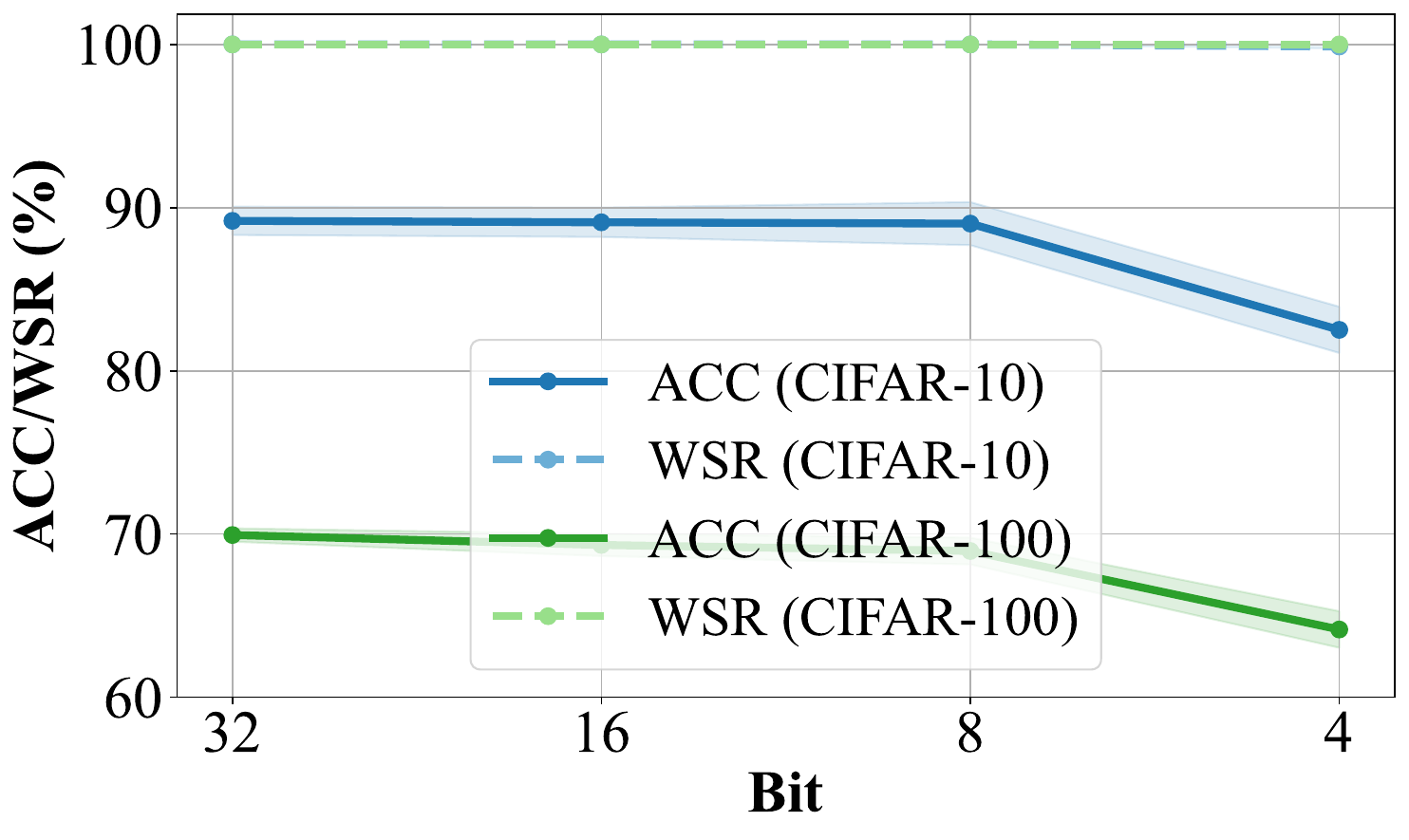}
        \subfloat{Weight Quantization}
        \label{subfig:target_quantization}
    \end{minipage}
    \hspace{0.1\linewidth} 
    \begin{minipage}{0.4\linewidth}
        \centering
        \includegraphics[width=\linewidth]{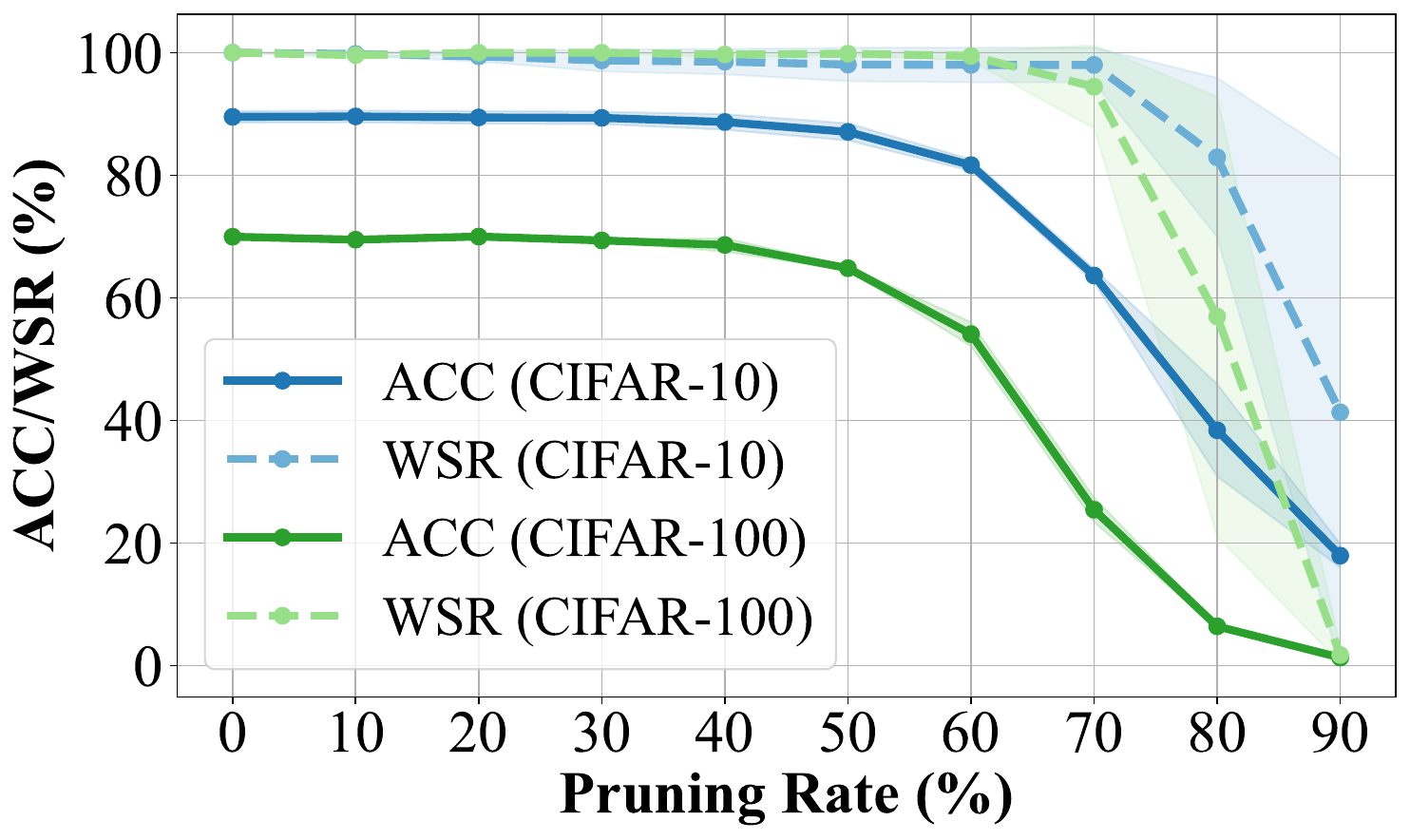}
        \subfloat{Pruning}
        \label{subfig:target_pruning}
    \end{minipage}

	\caption{Results of target model after model compression. (a) results with different levels of weight quantization. (b) results with varying pruning rates.}
	\label{fig:target_quantization_pruning} 
\end{figure*}

\textbf{Distillation.}
Distillation was initially proposed for model compression by cross-model knowledge transfer from a large pre-trained teacher model to a compact student model. We follow the classic setting to use the labeled training data to distill \textit{T2S} watermarked ResNet-18 target models to MobileNet-v2 \cite{mobilenetv2} student models. In addition to the the soft labels returned by the target model, the smaller model is also trained with the ground-truth labels of the training data. 

Table \ref{tab:distill_compression} shows the performance of the resulting compact models trained by distilling from different models watermarked with three approaches. Despite the classification performance remaining largely unchanged or even slightly improving after distillation due to the similar capacities of the teacher and student models, the watermarks embedded using the three different approaches have experienced significant degradation.  Compared with the results of \textit{Knockoff}, it is known that enhancing on knowledge on the classification task by learning from the ground-truth labels compromises watermarking. Nevertheless, the watermarks learned by \textit{T2S} demonstrate greater integrity, showing improvements of over 13\% on CIFAR-10 and 35\% on CIFAR-100 in comparison to \textit{SSW-S}.

\textbf{Quantization.}
We employed a standard weight quantization approach to change model weights from high precision (e.g., 32-bit) to lower precision (e.g., 8-bit or 4-bit).
As shown in Fig. \ref{fig:target_quantization_pruning}, our method demonstrates high robustness against quantization. The WSR remains at 100\% even under 4-bit quantization, whereas the ACC drops by more than 5\%.

\textbf{Pruning.}
We evaluated the changes in ACC and WSR under varying pruning rates (the proportion of weights removed), as shown in Fig.\ref{subfig:target_pruning}. At low pruning rates ($<$ 50\%), both the classification performance and watermark success rate are nearly unaffected. However, when the pruning rate exceeds 50\%, the classification accuracy begins to decline sharply, while the watermark success rate remains and drops significantly when the pruning rate surpasses 70\%. It indicates that the weights responsible for recognizing the watermark are less likely to be pruned than those related to the primary task. As a result, the watermark remains largely intact under practically acceptable pruning levels.

\subsection{Analytical studies\label{sec:ablation}}
Now we provide analytical studies with respect to the effectiveness of the fine-tuning phase in $T2S$, the impact of coefficient $\alpha$, and the selection of source and target classes on robustness against model stealing.
\subsubsection{Ablation of fine-tuning}
 To evaluate the effectiveness of the fine-tuning phase, we implement $T2S$ with the default setting (pre-train + fine-tuning) and the ``w/o rehearsal'' setting by ablating rehearsal-based fine-tuning. While feature-based trigger set is used by default, \textit{T2S} also works with other types of trigger sets. We apply our watermark tuning framework to
two other types of trigger sets called \textit{OOD} and \textit{Mix}. For OOD, we uses out-of-distribution images as the triggering inputs and labels them to a target class as in \cite{jia21}. For \textit{Mix}, we combines training images from two different source classes and labels them to the third class like in \cite{lv24}.

As shown in Table \ref{tab:ablation}, the proposed approach maintains consistent performance across different trigger sets with the default setting, demonstrating the general applicability of the proposed fine-tuning approach.
Watermarks embedded in a conventional method are largely removed for the three trigger types, but remain well-preserved when rehearsal-based fine-tuning is utilized. This demonstrates that the core robustness originates from the novel rehearsal-based watermark tuning, while feature-based triggers serve as an optional enhancement rather than a necessity.


\subsubsection{Impact of $\alpha$}\label{sec:alpha}

\begin{figure*}[htb]
\vskip 0.1in
	\centering
	\subfloat[Target Model ACC]{
        \includegraphics[width=0.23\linewidth]{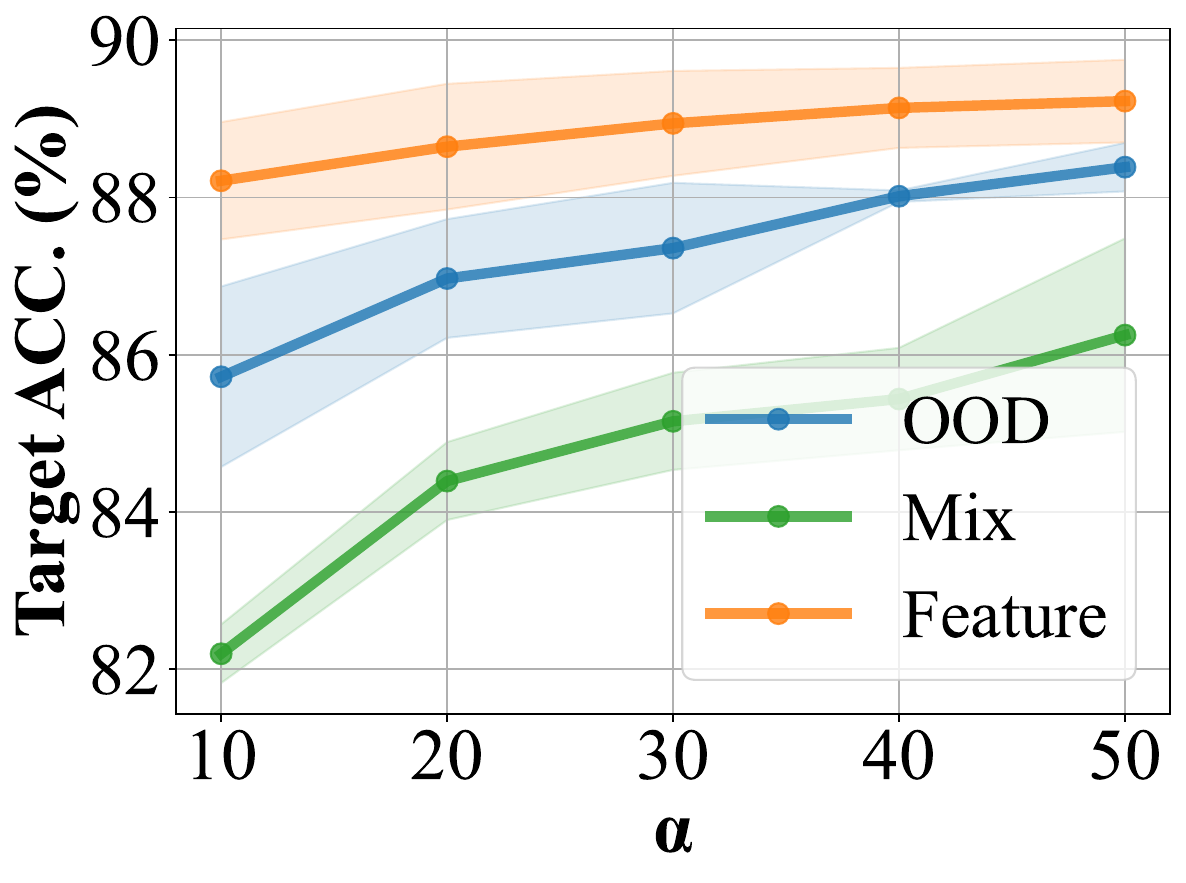}
    }
    \subfloat[Target Model WSR]{
        \includegraphics[width=0.23\linewidth]{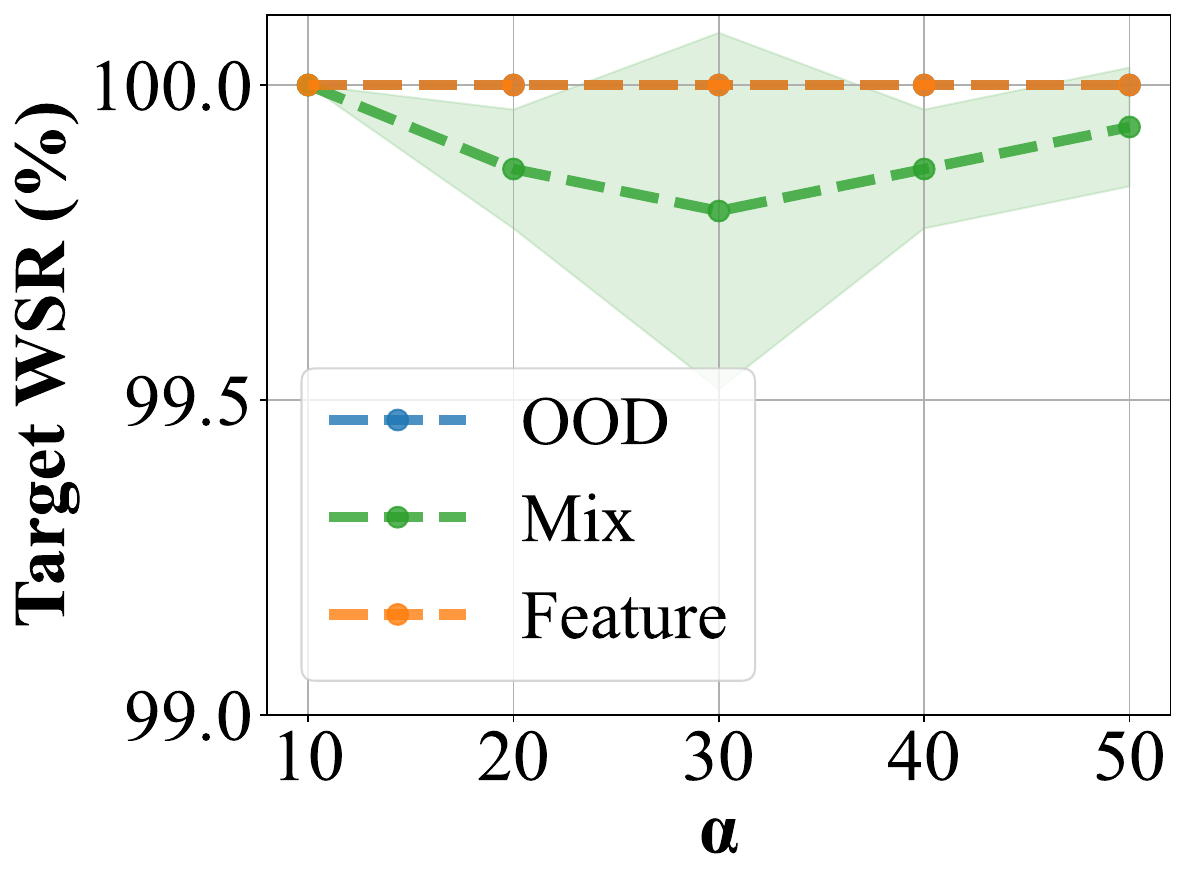}
        \label{subfig:target_model}
    }
    \subfloat[Stolen Model ACC]{
        \includegraphics[width=0.23\linewidth]{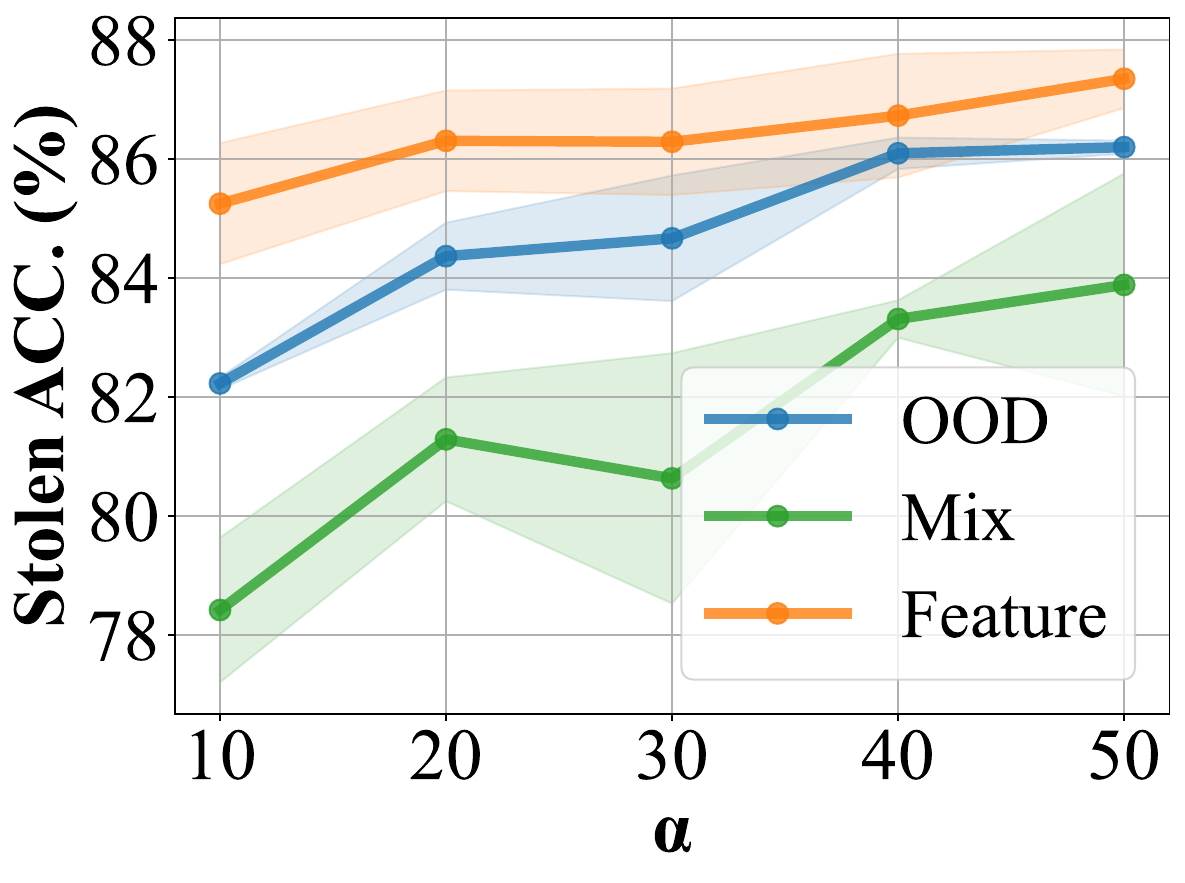}
        \label{subfig:target_model}
    }
    \subfloat[Stolen Model WSR]{
        \includegraphics[width=0.23\linewidth]{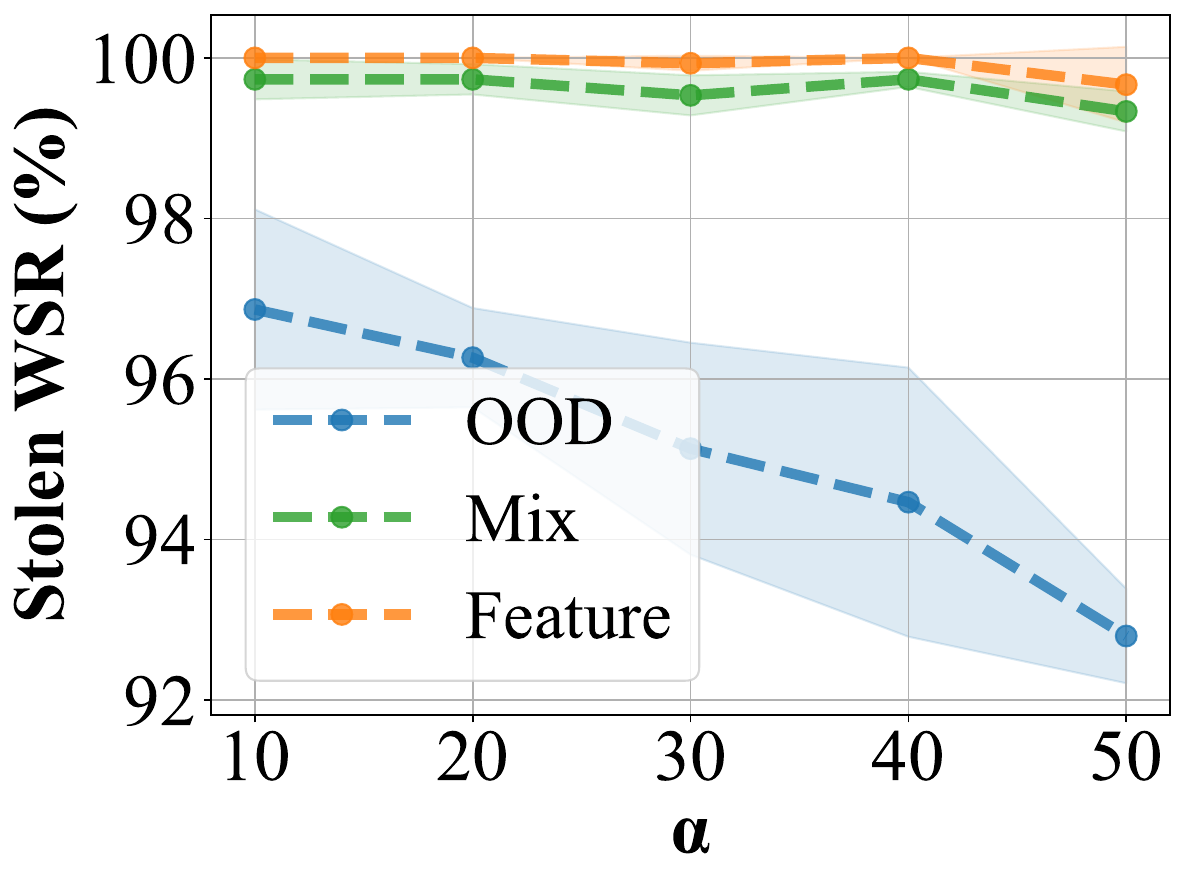}
        \label{subfig:target_model}
    }
	\caption{\textbf{Impact of $\alpha$ in \textit{T2S} on CIFAR-10.} Target model ACC and stolen model WSR (\%) with three trigger sets. Stolen models are extracted with \textit{Knockoff} (soft label).}
	\label{fig:alpha_cifar10} 
\vskip -0.1in
\end{figure*}

\begin{figure*}[htb]
\vskip 0.1in
	\centering
	\subfloat[Target Model ACC]{
        \includegraphics[width=0.23\linewidth]{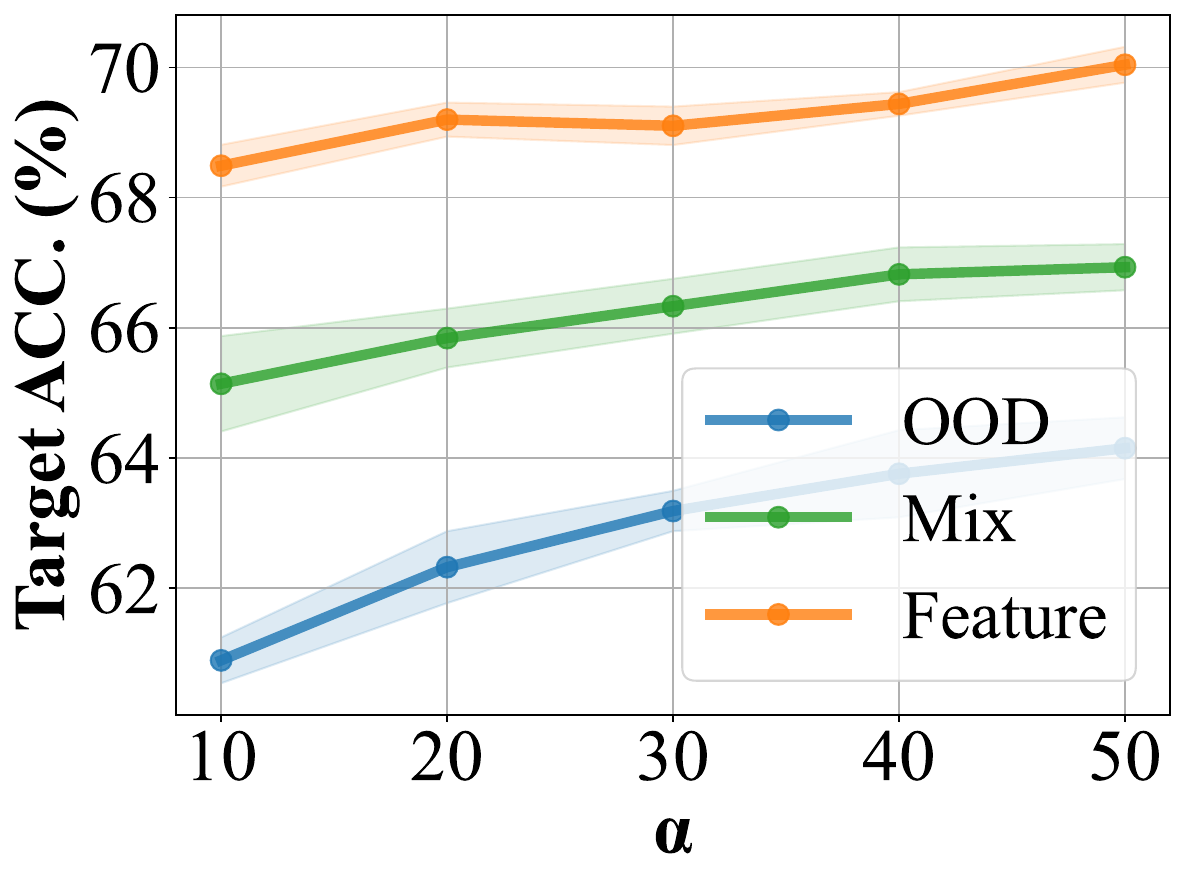}
    }
    \subfloat[Target Model WSR]{
        \includegraphics[width=0.23\linewidth]{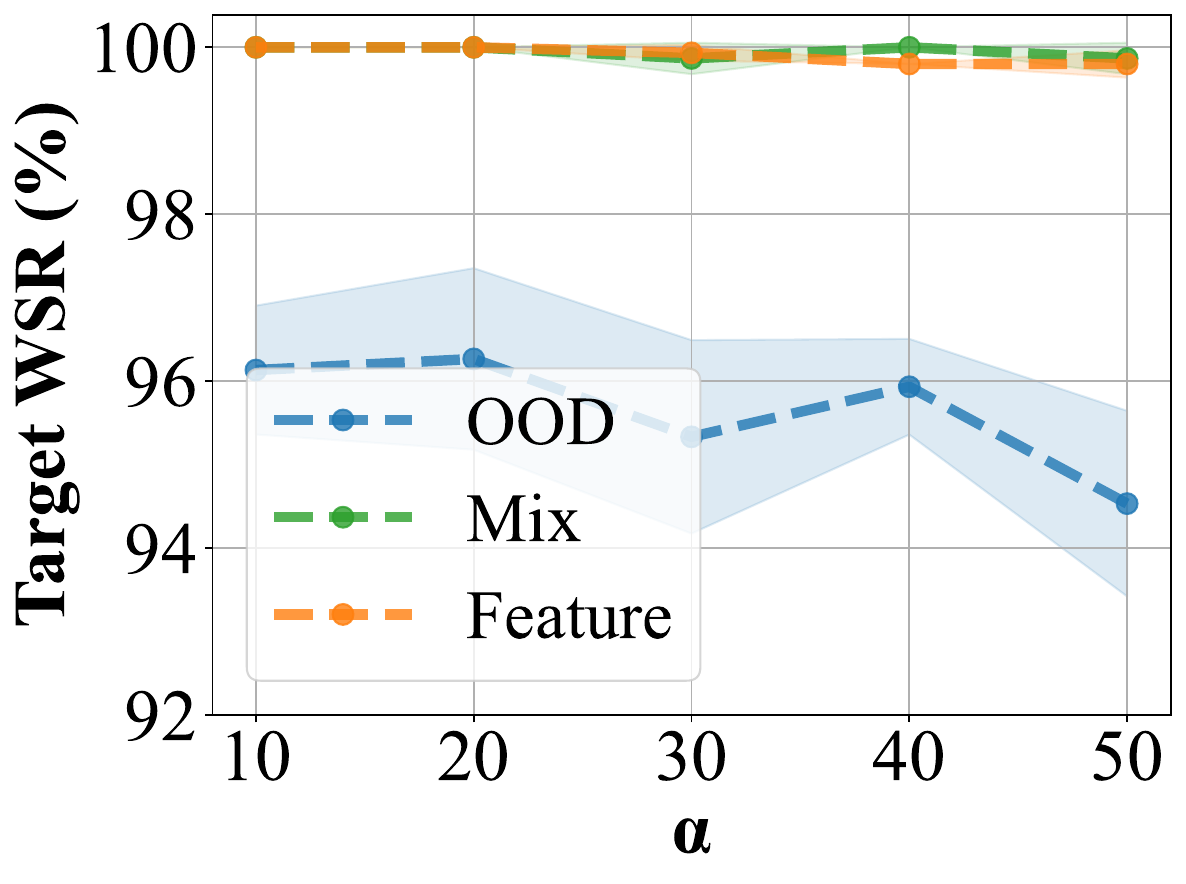}
        \label{subfig:target_model}
    }
    \subfloat[Stolen Model ACC]{
        \includegraphics[width=0.23\linewidth]{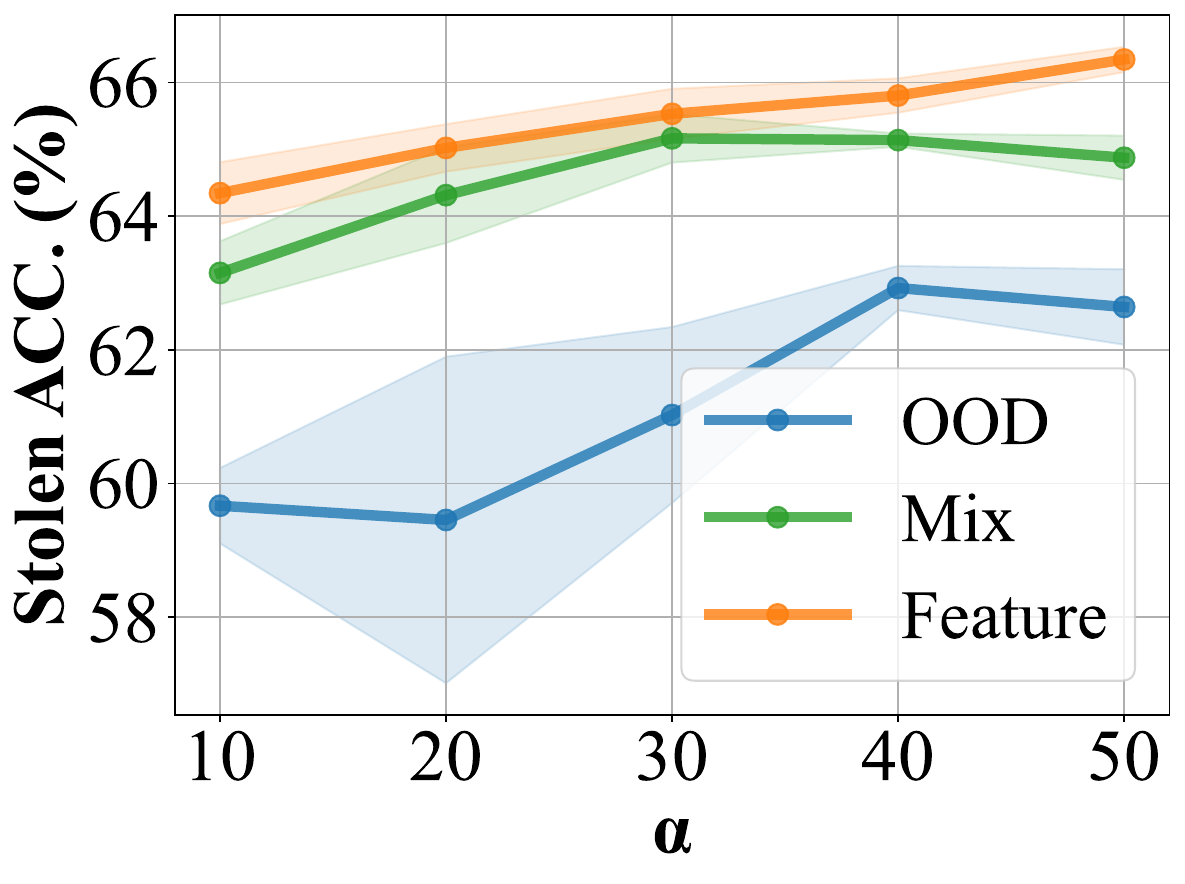}
        \label{subfig:target_model}
    }
    \subfloat[Stolen Model WSR]{
        \includegraphics[width=0.23\linewidth]{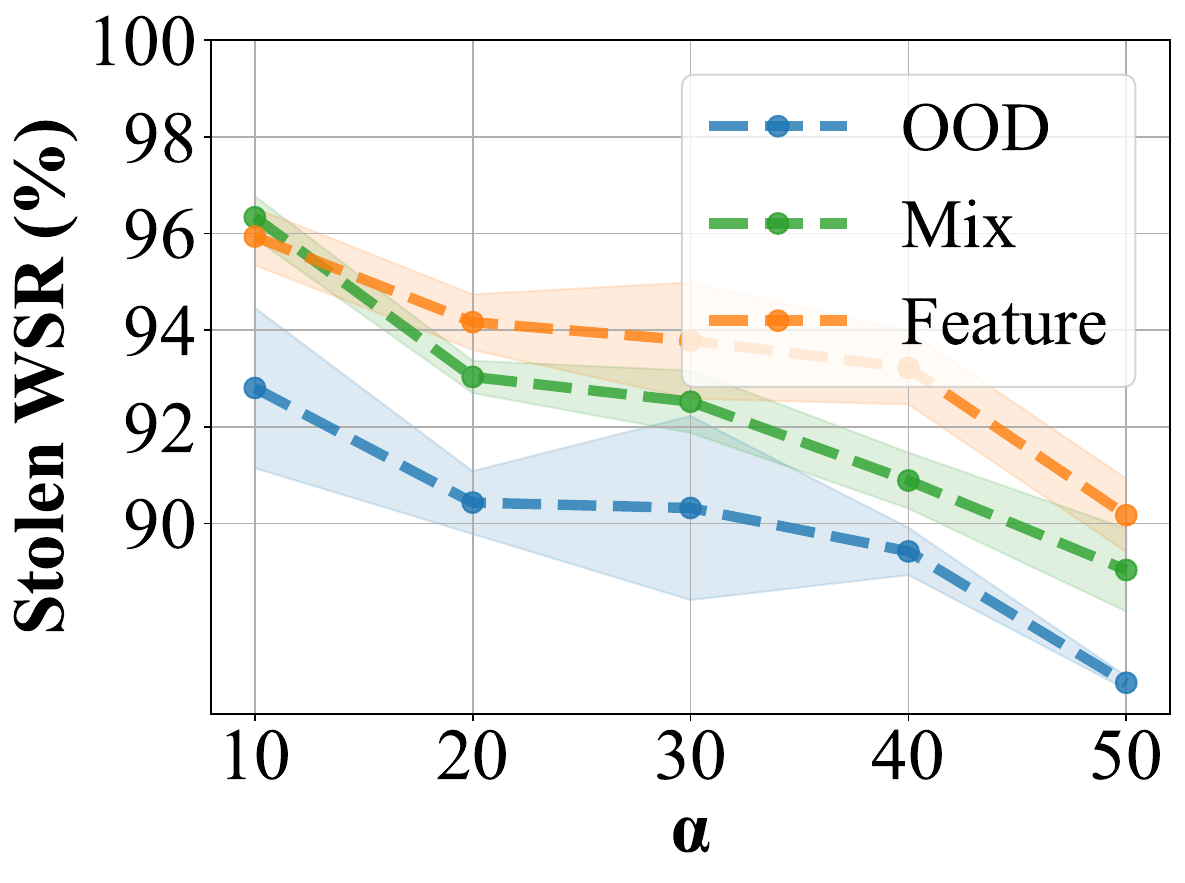}
        \label{subfig:target_model}
    }
	\caption{\textbf{Impact of $\alpha$ in \textit{T2S} on CIFAR-100.} Target model ACC and stolen model WSR (\%)  with three trigger sets. Stolen models are extracted with \textit{Knockoff} (soft label).}
	\label{fig:alpha_cifar100} 
\vskip -0.1in
\end{figure*}

 We now analyze the influence of $\alpha$ in Eq.(\ref{eq:final}), where a larger $\alpha$ increases the weight on model utility while reducing the priority of watermark retention. Results in
 Fig.\ref{fig:alpha_cifar10} and  Fig.\ref{fig:alpha_cifar100}  confirm that smaller values of $\alpha$ achieve better watermark performance at the cost of some model accuracy. It is also observed that feature-based trigger set is more favorable than the other two types of trigger sets on the two CIFAR datasets. This superiority is demonstrated by its ability to maintain high ACCs and WSRs across all values of $\alpha$.

\begin{table}[htbp]
\centering
\caption{\textbf{Ablation study.} ACC and WSR (\%) of \textit{T2S} with feature-based, OOD, and Mix trigger sets. Stolen models: ResNet-18, \textit{Knockoff} (soft label). ``w/o rehearsal'' indicates no fine-tuning.} \label{tab:ablation}
\resizebox{1\linewidth}{!}{  
\begin{tabular}{ccccc}
\toprule
\multirow{2}{*}{\textbf{Method}} & \multicolumn{2}{c}{\textbf{CIFAR-10}} & \multicolumn{2}{c}{\textbf{CIFAR-100}}\\
\cmidrule(r){2-3}\cmidrule{4-5} & ACC & WSR & ACC & WSR \\
\midrule
\textbf{default}\\\midrule
     OOD            & 86.10$\pm$0.33  & 94.47$\pm$2.05  & 62.92$\pm$0.41  & 94.60$\pm$1.00 \\
     Mix            & 83.31$\pm$0.39  & 99.73$\pm$0.12  & 65.14$\pm$0.12  & 94.67$\pm$2.34 \\
     feature  & 86.87$\pm$0.55  & 99.87$\pm$0.23  & 65.80$\pm$0.32  & 97.40$\pm$2.78 \\
\midrule
\textbf{w/o rehearsal}\\\midrule
     OOD            & 86.04$\pm$0.96  & 52.33$\pm$6.66  & 66.21$\pm$0.68  & 52.67$\pm$10.50 \\
     Mix            & 86.69$\pm$0.38  & 31.30$\pm$14.84 & 66.39$\pm$0.58  & 21.53$\pm$14.44 \\
     feature  & 86.97$\pm$0.48  & 3.73$\pm$2.58   & 66.28$\pm$0.09  & 28.87$\pm$17.19 \\
\bottomrule
\end{tabular}
}
\end{table}

\subsubsection{Impact of source and target class selection.}\label{sec:triggertype}
\begin{figure*}[!t]
    \centering
    \subfloat[Target model ACC]{%
        \includegraphics[width=0.45\textwidth]{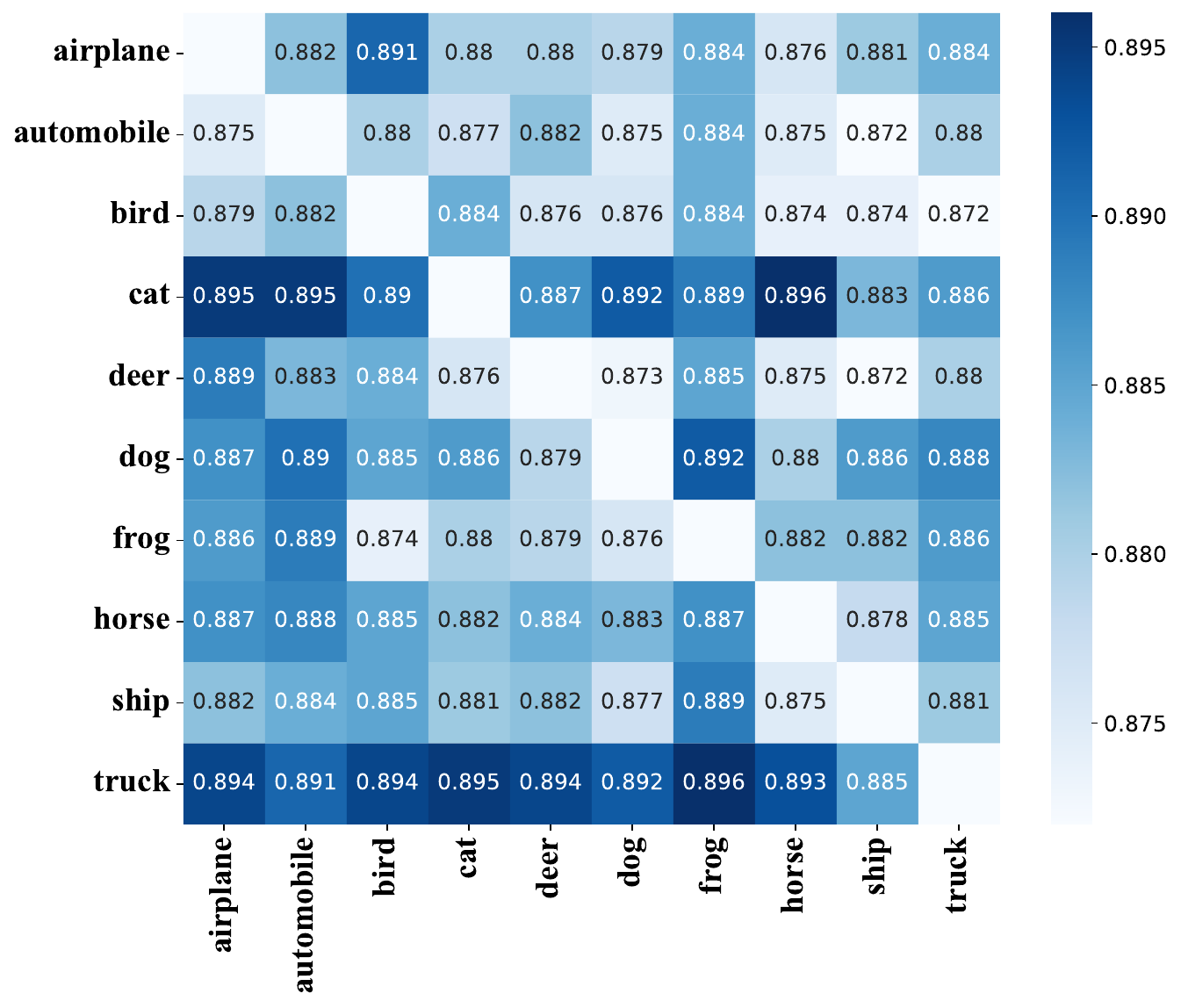}
    }
    \hspace{0.01\linewidth} 
    \subfloat[Target model WSR]{%
        \includegraphics[width=0.45\textwidth]{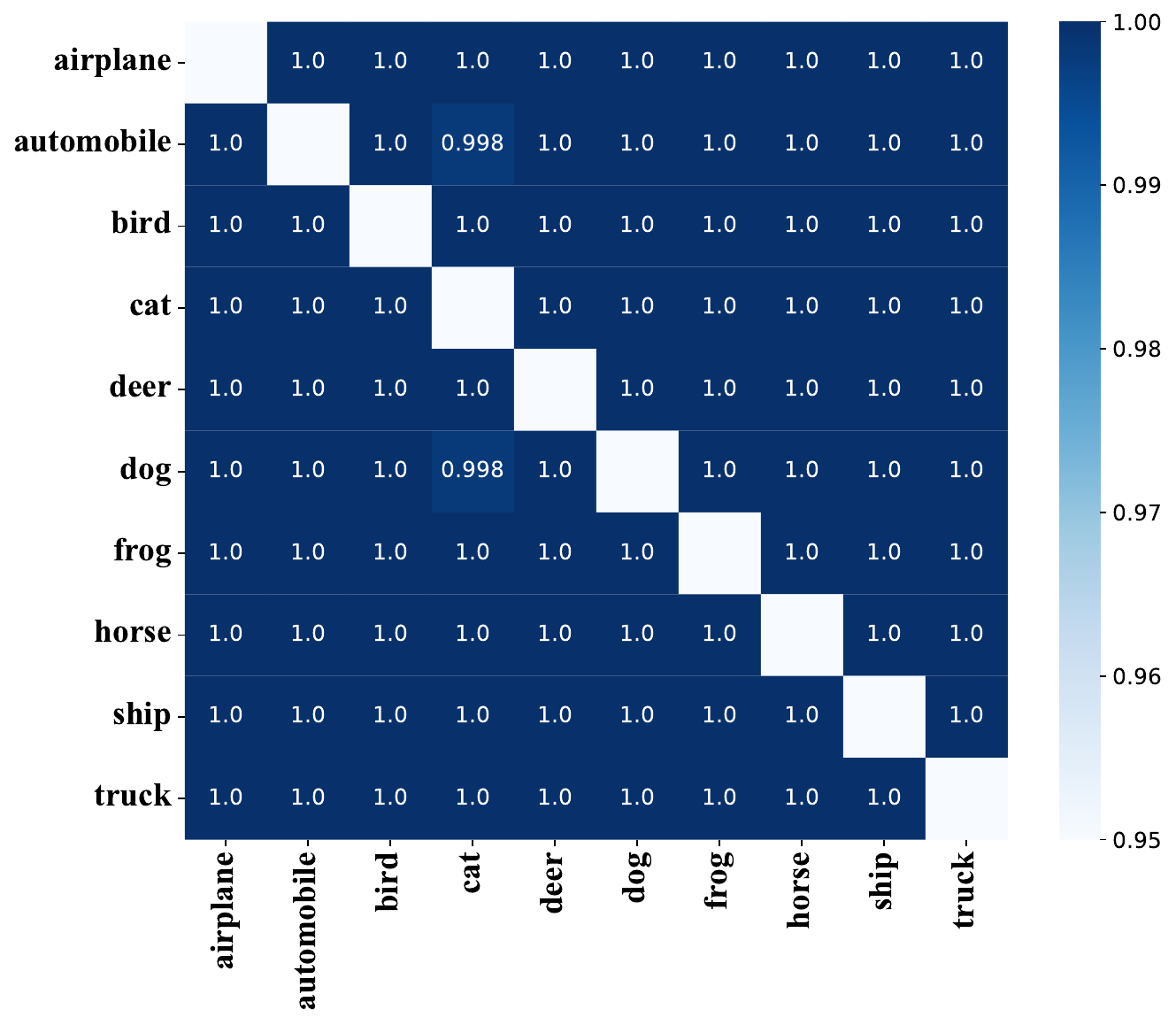}
    }
    \caption{Heatmap of \textbf{Target model} ACC and WSRs with respect to different source–target pairs of CIFAR-10.}
    \label{fig:source_target_metrics1}
\end{figure*}

\begin{figure*}[!t]
    \centering
    \subfloat[Stolen model ACC]{%
        \includegraphics[width=0.45\textwidth]{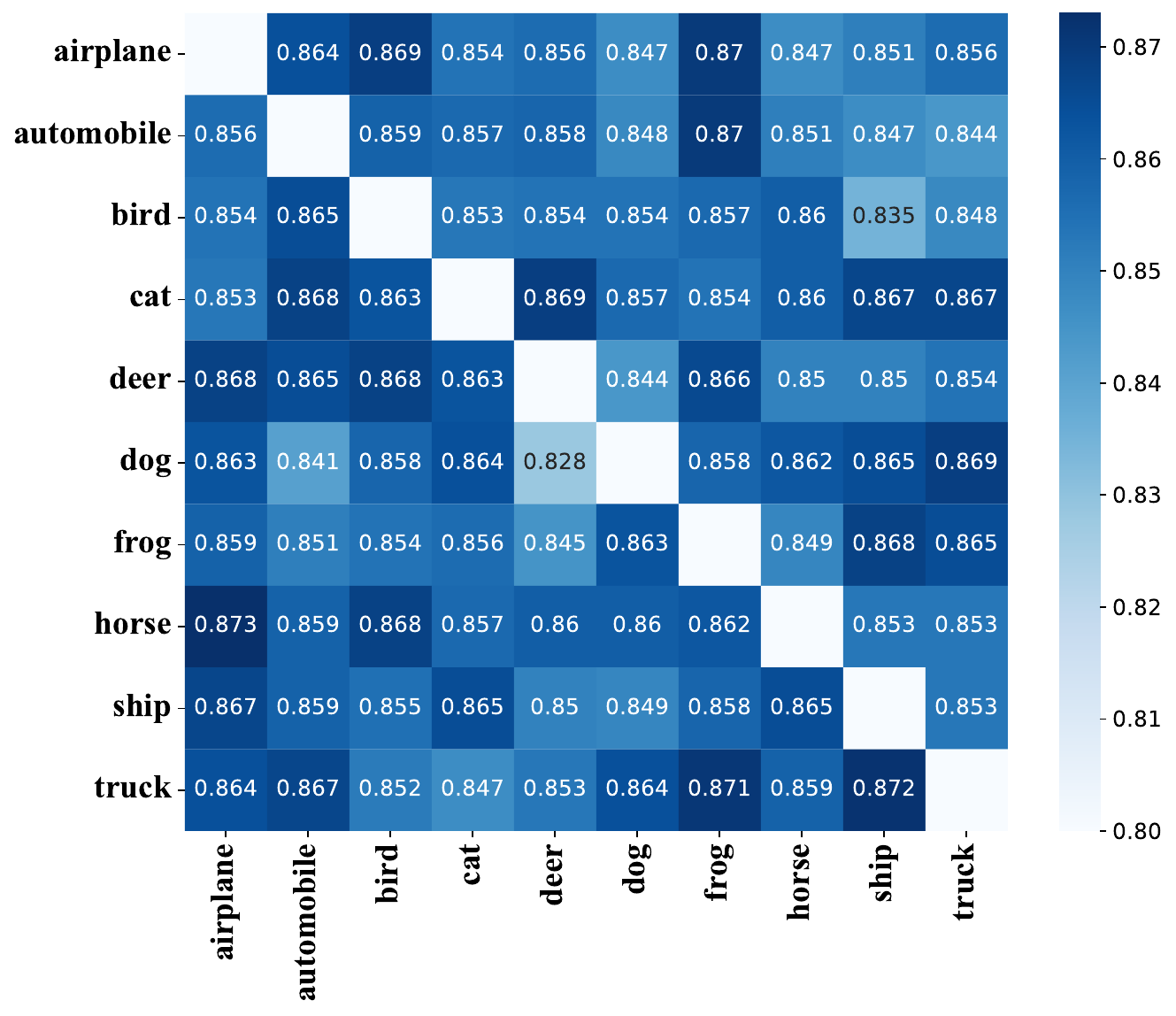}
    }
    \hspace{0.01\linewidth} 
    \subfloat[Stolen model WSR]{%
        \includegraphics[width=0.45\textwidth]{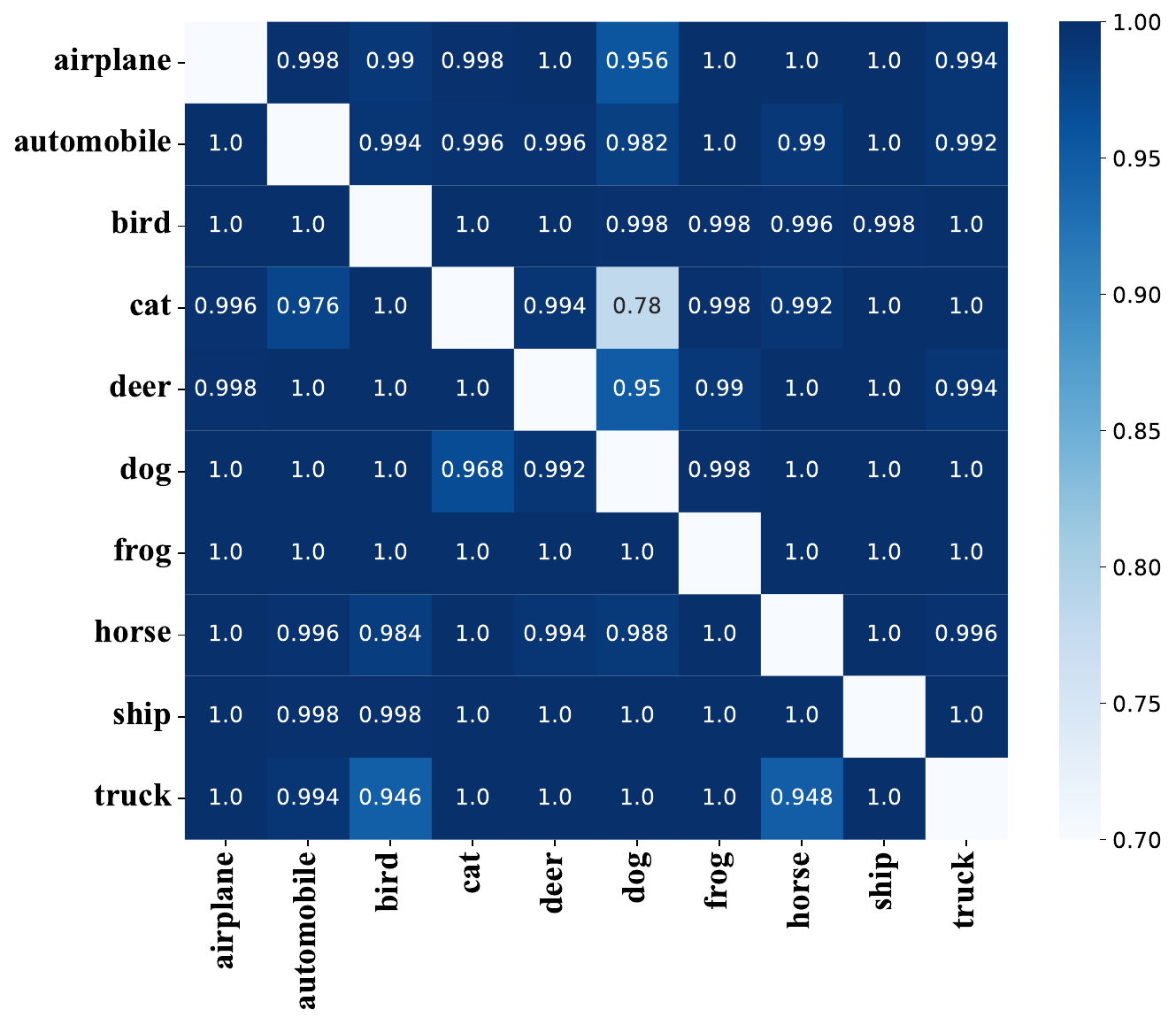}
    }
    \caption{Heatmap of \textbf{Stolen model} ACC and WSRs with respect to different source–target pairs of CIFAR-10.}
    \label{fig:source_target_metrics2}
\end{figure*}

We obtained the results of the target and stolen model with all combinations of source and target classes to obtain feature-based trigger set for CIFAR-10. Specifically, the target model ACCs and WSRs are given in Fig.\ref{fig:source_target_metrics1} and the stolen model ACCs and WSRs in
 Fig.\ref{fig:source_target_metrics2}. Stolen models are extracted with \textit{Knockoff} (soft label) from \textit{T2S} watermarked models.

  The average target model ACC (\%) over all the combinations is $88.34\pm 0.63$.
 We observed the lowest WSR of 78\% when using ``cat'' as the source class and ``dog'' as the target class. 
 The average WSR (\%) of the stolen model over all these combinations is $99.27\pm 2.51$.
  These results show that while degradation occurs in final performance when the source-target pair is not selected properly, \textit{T2S} can achieve reasonable utility and protection in most cases. 

The average value of target model WSR (\%) is  $100 \pm 0.03$, indicating that the choice of source-target pairs has a minimal impact on the \textit{T2S} watermarked models. The average value of stolen model ACC (\%) is  $85.78 \pm 0.85$, showing that the performance of the stolen model is not very sensitive to the selection of the source-target classes in watermarking.

\section{Limitation}
While simulation has been widely employed in developing both attack and defense techniques, it indeed makes training more expensive. 

In the proposed approach, the simulation is exclusively performed during the fine-tuning phase (typically completed within 1-2 epochs) rather than the full pre-training stage. This phased implementation helps to minimize the simulation caused training overhead and also enables us to apply potential advanced fine-tuning techniques with better computation and data efficiency, e.g., using a small subset for fine-tuning, enhancing the scalability of our approach to large datasets.

Our watermark fine-tuning process requires computing second-order derivatives, which imposes significant GPU memory demands. In our vanilla implementation (tested on a single RTX 3090 with 24GB VRAM), fine-tuning ResNet-18 for one epoch on CIFAR-10 with a batch size of $B_D=100$ requires 425 seconds and 4.9GB GPU memory. Scaling up the batch size to $B_D=1000$ increases GPU memory usage to 17.6GB while reducing training time to 137 seconds per epoch. While our approach places relatively high demands on GPU memory, it remains particularly well-suited for protection of edge-device deployed models. Techniques such as gradient checkpointing and distributed training across multiple GPUs may further be employed to enhance scalability for larger models.



\section{Conclusion}
\label{sec:conclude}
We worked on model watermarking for robust protection against extraction-based black-box model stealing attacks. A novel watermark fine-tuning framework is proposed to fully leverage extraction-simulation by updating the initially watermarked model directly with the feedback of a simulated stolen model. The proposed method does not require to collect individually trained models to ensure low false positives and is applicable to various trigger set designs. Extensive experimental evaluation demonstrates that our method significantly enhances watermark robustness against diverse removal techniques with negligible impact on model utility. Notably, it greatly improves the retention of watermark integrity upon model extraction, making it a promising post-defense against model theft.
While our watermark fine-tuning process demands higher GPU memory due to second-order derivative computations, it remains highly effective for watermarking edge-deployed AI models with moderate parameter scales. This makes it critical for securing smart city intelligent edge, industrial IoT, and healthcare edge devices.



\bibliographystyle{unsrt}
\bibliography{bare_jrnl_new_sample4}

\vfill

\end{document}